\ifx\newheadisloaded\relax\immediate\write16{***already loaded}\endinput\else\let\newheadisloaded=\relax\fi
\gdef\islinuxolivetti{F}
\gdef\PSfonts{T}
\magnification\magstep1

\newdimen\papwidth
\newdimen\papheight
\newskip\beforesectionskipamount  
\newskip\sectionskipamount 
\def\sectionskip{\vskip\sectionskipamount}
\def\beforesectionskip{\vskip\beforesectionskipamount}
\papwidth=16truecm
\if F\islinuxolivetti
\papheight=22truecm
\voffset=0.4truecm
\hoffset=0.4truecm
\else
\papheight=16truecm
\voffset=-1.5truecm
\hoffset=0.4truecm
\fi
\hsize=\papwidth
\vsize=\papheight
\nopagenumbers
\headline={\ifnum\pageno>1 {\hss\tenrm-\ \folio\ -\hss} \else
{\hfill}\fi}
\newdimen\texpscorrection
\texpscorrection=0.15truecm 

\def\sectionsize{\twelvepoint}
\def\sectiontype{\bf}
\def\subsectionsize{}
\def\subsectiontype{\bf}
\def\em{\sl}
\newfam\truecmsy
\newfam\truecmr
\newfam\msbfam
\newfam\scriptfam
\newfam\truecmsy
\newskip\ttglue 
\if T\islinuxolivetti
\papheight=11.5truecm
\fi
\if F\PSfonts
\font\twelverm=cmr12
\font\tenrm=cmr10
\font\eightrm=cmr8
\font\sevenrm=cmr7
\font\sixrm=cmr6
\font\fiverm=cmr5

\font\twelvebf=cmbx12
\font\tenbf=cmbx10
\font\eightbf=cmbx8
\font\sevenbf=cmbx7
\font\sixbf=cmbx6
\font\fivebf=cmbx5

\font\twelveit=cmti12
\font\tenit=cmti10
\font\eightit=cmti8
\font\sevenit=cmti7
\font\sixit=cmti6
\font\fiveit=cmti5

\font\twelvesl=cmsl12
\font\tensl=cmsl10
\font\eightsl=cmsl8
\font\sevensl=cmsl7
\font\sixsl=cmsl6
\font\fivesl=cmsl5

\font\twelvei=cmmi12
\font\teni=cmmi10
\font\eighti=cmmi8
\font\seveni=cmmi7
\font\sixi=cmmi6
\font\fivei=cmmi5

\font\twelvesy=cmsy10	at	12pt
\font\tensy=cmsy10
\font\eightsy=cmsy8
\font\sevensy=cmsy7
\font\sixsy=cmsy6
\font\fivesy=cmsy5
\font\twelvetruecmsy=cmsy10	at	12pt
\font\tentruecmsy=cmsy10
\font\eighttruecmsy=cmsy8
\font\seventruecmsy=cmsy7
\font\sixtruecmsy=cmsy6
\font\fivetruecmsy=cmsy5

\font\twelvetruecmr=cmr12
\font\tentruecmr=cmr10
\font\eighttruecmr=cmr8
\font\seventruecmr=cmr7
\font\sixtruecmr=cmr6
\font\fivetruecmr=cmr5

\font\twelvebf=cmbx12
\font\tenbf=cmbx10
\font\eightbf=cmbx8
\font\sevenbf=cmbx7
\font\sixbf=cmbx6
\font\fivebf=cmbx5

\font\twelvett=cmtt12
\font\tentt=cmtt10
\font\eighttt=cmtt8

\font\twelveex=cmex10	at	12pt
\font\tenex=cmex10

\font\twelvemsb=msbm10	at	12pt
\font\tenmsb=msbm10
\font\eightmsb=msbm8
\font\sevenmsb=msbm7
\font\sixmsb=msbm6
\font\fivemsb=msbm5

\font\twelvescr=eusm10 at 12pt
\font\tenscr=eusm10
\font\eightscr=eusm8
\font\sevenscr=eusm7
\font\sixscr=eusm6
\font\fivescr=eusm5
\fi
\if T\PSfonts
\font\twelverm=ptmr	at	12pt
\font\tenrm=ptmr	at	10pt
\font\eightrm=ptmr	at	8pt
\font\sevenrm=ptmr	at	7pt
\font\sixrm=ptmr	at	6pt
\font\fiverm=ptmr	at	5pt

\font\twelvebf=ptmb	at	12pt
\font\tenbf=ptmb	at	10pt
\font\eightbf=ptmb	at	8pt
\font\sevenbf=ptmb	at	7pt
\font\sixbf=ptmb	at	6pt
\font\fivebf=ptmb	at	5pt

\font\twelveit=ptmri	at	12pt
\font\tenit=ptmri	at	10pt
\font\eightit=ptmri	at	8pt
\font\sevenit=ptmri	at	7pt
\font\sixit=ptmri	at	6pt
\font\fiveit=ptmri	at	5pt

\font\twelvesl=ptmro	at	12pt
\font\tensl=ptmro	at	10pt
\font\eightsl=ptmro	at	8pt
\font\sevensl=ptmro	at	7pt
\font\sixsl=ptmro	at	6pt
\font\fivesl=ptmro	at	5pt

\font\twelvei=cmmi12
\font\teni=cmmi10
\font\eighti=cmmi8
\font\seveni=cmmi7
\font\sixi=cmmi6
\font\fivei=cmmi5

\font\twelvesy=cmsy10	at	12pt
\font\tensy=cmsy10
\font\eightsy=cmsy8
\font\sevensy=cmsy7
\font\sixsy=cmsy6
\font\fivesy=cmsy5
\font\twelvetruecmsy=cmsy10	at	12pt
\font\tentruecmsy=cmsy10
\font\eighttruecmsy=cmsy8
\font\seventruecmsy=cmsy7
\font\sixtruecmsy=cmsy6
\font\fivetruecmsy=cmsy5

\font\twelvetruecmr=cmr12
\font\tentruecmr=cmr10
\font\eighttruecmr=cmr8
\font\seventruecmr=cmr7
\font\sixtruecmr=cmr6
\font\fivetruecmr=cmr5

\font\twelvebf=cmbx12
\font\tenbf=cmbx10
\font\eightbf=cmbx8
\font\sevenbf=cmbx7
\font\sixbf=cmbx6
\font\fivebf=cmbx5

\font\twelvett=cmtt12
\font\tentt=cmtt10
\font\eighttt=cmtt8

\font\twelveex=cmex10	at	12pt
\font\tenex=cmex10

\font\twelvemsb=msbm10	at	12pt
\font\tenmsb=msbm10
\font\eightmsb=msbm8
\font\sevenmsb=msbm7
\font\sixmsb=msbm6
\font\fivemsb=msbm5

\font\twelvescr=eusm10 at 12pt
\font\tenscr=eusm10
\font\eightscr=eusm8
\font\sevenscr=eusm7
\font\sixscr=eusm6
\font\fivescr=eusm5
\fi
\def\eightpoint{\def\rm{\fam0\eightrm}%
\textfont0=\eightrm
  \scriptfont0=\sixrm
  \scriptscriptfont0=\fiverm 
\textfont1=\eighti
  \scriptfont1=\sixi
  \scriptscriptfont1=\fivei 
\textfont2=\eightsy
  \scriptfont2=\sixsy
  \scriptscriptfont2=\fivesy 
\textfont3=\tenex
  \scriptfont3=\tenex
  \scriptscriptfont3=\tenex 
\textfont\itfam=\eightit
  \scriptfont\itfam=\sixit
  \scriptscriptfont\itfam=\fiveit 
  \def\it{\fam\itfam\eightit}%
\textfont\slfam=\eightsl
  \scriptfont\slfam=\sixsl
  \scriptscriptfont\slfam=\fivesl 
  \def\sl{\fam\slfam\eightsl}%
\textfont\ttfam=\eighttt
  \def\tt{\fam\ttfam\eighttt}%
\textfont\bffam=\eightbf
  \scriptfont\bffam=\sixbf
  \scriptscriptfont\bffam=\fivebf
  \def\bf{\fam\bffam\eightbf}%
\textfont\scriptfam=\eightscr
  \scriptfont\scriptfam=\sixscr
  \scriptscriptfont\scriptfam=\fivescr
  \def\script{\fam\scriptfam\eightscr}%
\textfont\msbfam=\eightmsb
  \scriptfont\msbfam=\sixmsb
  \scriptscriptfont\msbfam=\fivemsb
  \def\bb{\fam\msbfam\eightmsb}%
\textfont\truecmr=\eighttruecmr
  \scriptfont\truecmr=\sixtruecmr
  \scriptscriptfont\truecmr=\fivetruecmr
  \def\truerm{\fam\truecmr\eighttruecmr}%
\textfont\truecmsy=\eighttruecmsy
  \scriptfont\truecmsy=\sixtruecmsy
  \scriptscriptfont\truecmsy=\fivetruecmsy
\tt \ttglue=.5em plus.25em minus.15em 
\normalbaselineskip=9pt
\setbox\strutbox=\hbox{\vrule height7pt depth2pt width0pt}%
\normalbaselines
\rm
}

\def\tenpoint{\def\rm{\fam0\tenrm}%
\textfont0=\tenrm
  \scriptfont0=\sevenrm
  \scriptscriptfont0=\fiverm 
\textfont1=\teni
  \scriptfont1=\seveni
  \scriptscriptfont1=\fivei 
\textfont2=\tensy
  \scriptfont2=\sevensy
  \scriptscriptfont2=\fivesy 
\textfont3=\tenex
  \scriptfont3=\tenex
  \scriptscriptfont3=\tenex 
\textfont\itfam=\tenit
  \scriptfont\itfam=\sevenit
  \scriptscriptfont\itfam=\fiveit 
  \def\it{\fam\itfam\tenit}%
\textfont\slfam=\tensl
  \scriptfont\slfam=\sevensl
  \scriptscriptfont\slfam=\fivesl 
  \def\sl{\fam\slfam\tensl}%
\textfont\ttfam=\tentt
  \def\tt{\fam\ttfam\tentt}%
\textfont\bffam=\tenbf
  \scriptfont\bffam=\sevenbf
  \scriptscriptfont\bffam=\fivebf
  \def\bf{\fam\bffam\tenbf}%
\textfont\scriptfam=\tenscr
  \scriptfont\scriptfam=\sevenscr
  \scriptscriptfont\scriptfam=\fivescr
  \def\script{\fam\scriptfam\tenscr}%
\textfont\msbfam=\tenmsb
  \scriptfont\msbfam=\sevenmsb
  \scriptscriptfont\msbfam=\fivemsb
  \def\bb{\fam\msbfam\tenmsb}%
\textfont\truecmr=\tentruecmr
  \scriptfont\truecmr=\seventruecmr
  \scriptscriptfont\truecmr=\fivetruecmr
  \def\truerm{\fam\truecmr\tentruecmr}%
\textfont\truecmsy=\tentruecmsy
  \scriptfont\truecmsy=\seventruecmsy
  \scriptscriptfont\truecmsy=\fivetruecmsy
\tt \ttglue=.5em plus.25em minus.15em 
\normalbaselineskip=12pt
\setbox\strutbox=\hbox{\vrule height8.5pt depth3.5pt width0pt}%
\normalbaselines
\rm
}

\def\twelvepoint{\def\rm{\fam0\twelverm}%
\textfont0=\twelverm
  \scriptfont0=\tenrm
  \scriptscriptfont0=\eightrm 
\textfont1=\twelvei
  \scriptfont1=\teni
  \scriptscriptfont1=\eighti 
\textfont2=\twelvesy
  \scriptfont2=\tensy
  \scriptscriptfont2=\eightsy 
\textfont3=\twelveex
  \scriptfont3=\twelveex
  \scriptscriptfont3=\twelveex 
\textfont\itfam=\twelveit
  \scriptfont\itfam=\tenit
  \scriptscriptfont\itfam=\eightit 
  \def\it{\fam\itfam\twelveit}%
\textfont\slfam=\twelvesl
  \scriptfont\slfam=\tensl
  \scriptscriptfont\slfam=\eightsl 
  \def\sl{\fam\slfam\twelvesl}%
\textfont\ttfam=\twelvett
  \def\tt{\fam\ttfam\twelvett}%
\textfont\bffam=\twelvebf
  \scriptfont\bffam=\tenbf
  \scriptscriptfont\bffam=\eightbf
  \def\bf{\fam\bffam\twelvebf}%
\textfont\scriptfam=\twelvescr
  \scriptfont\scriptfam=\tenscr
  \scriptscriptfont\scriptfam=\eightscr
  \def\script{\fam\scriptfam\twelvescr}%
\textfont\msbfam=\twelvemsb
  \scriptfont\msbfam=\tenmsb
  \scriptscriptfont\msbfam=\eightmsb
  \def\bb{\fam\msbfam\twelvemsb}%
\textfont\truecmr=\twelvetruecmr
  \scriptfont\truecmr=\tentruecmr
  \scriptscriptfont\truecmr=\eighttruecmr
  \def\truerm{\fam\truecmr\twelvetruecmr}%
\textfont\truecmsy=\twelvetruecmsy
  \scriptfont\truecmsy=\tentruecmsy
  \scriptscriptfont\truecmsy=\eighttruecmsy
\tt \ttglue=.5em plus.25em minus.15em 
\setbox\strutbox=\hbox{\vrule height7pt depth2pt width0pt}%
\normalbaselineskip=15pt
\normalbaselines
\rm
}
%
\fontdimen16\tensy=2.7pt
\fontdimen13\tensy=4.3pt
\fontdimen17\tensy=2.7pt
\fontdimen14\tensy=4.3pt
\fontdimen18\tensy=4.3pt
\fontdimen16\eightsy=2.7pt
\fontdimen13\eightsy=4.3pt
\fontdimen17\eightsy=2.7pt
\fontdimen14\eightsy=4.3pt
\fontdimen18\sevensy=4.3pt
\fontdimen16\sevensy=1.8pt
\fontdimen13\sevensy=4.3pt
\fontdimen17\sevensy=2.7pt
\fontdimen14\sevensy=4.3pt
\fontdimen18\sevensy=4.3pt
%
\def\hexnumber#1{\ifcase#1 0\or1\or2\or3\or4\or5\or6\or7\or8\or9\or
 A\or B\or C\or D\or E\or F\fi}
\mathcode`\=="3\hexnumber\truecmr3D
\mathchardef\not="3\hexnumber\truecmsy36
\mathcode`\+="2\hexnumber\truecmr2B
\mathcode`\(="4\hexnumber\truecmr28
\mathcode`\)="5\hexnumber\truecmr29
\mathcode`\!="5\hexnumber\truecmr21
\mathcode`\(="4\hexnumber\truecmr28
\mathcode`\)="5\hexnumber\truecmr29

\def\bar{\mathaccent"0\hexnumber\truecmr16 }

\def\hat{\mathaccent"0\hexnumber\truecmr5E }
\def\dot{\mathaccent"0\hexnumber\truecmr5F }
\def\Phi{\mathchar"0\hexnumber\truecmr08 }
\def\Gamma {\mathchar"0\hexnumber\truecmr00 }
\def\Delta {\mathchar"0\hexnumber\truecmr01 }
\def\Theta {\mathchar"0\hexnumber\truecmr02 }
\def\Lambda{\mathchar"0\hexnumber\truecmr03 }
\def\Xi {\mathchar"0\hexnumber\truecmr04 }
\def\Pi{\mathchar"0\hexnumber\truecmr05 }
\def\Sigma{\mathchar"0\hexnumber\truecmr06 }
\def\Upsilon {\mathchar"0\hexnumber\truecmr07 }
\def\Phi {\mathchar"0\hexnumber\truecmr08 }
\def\Psi {\mathchar"0\hexnumber\truecmr09 }
\def\Omega{\mathchar"0\hexnumber\truecmr0A }
\newcount\EQNcount \EQNcount=1
\newcount\CLAIMcount \CLAIMcount=1
\newcount\SECTIONcount \SECTIONcount=0
\newcount\SUBSECTIONcount \SUBSECTIONcount=1
\def\ifff(#1,#2,#3){\ifundefined{#1#2}%
\expandafter\xdef\csname #1#2\endcsname{#3}\else%
\fi}
\def\NEWDEF #1,#2,#3 {\ifff({#1},{#2},{#3})}
\def\actualnumber{\number\SECTIONcount}
\def\EQ(#1){\lmargin(#1)\eqno\tag(#1)}
\def\NR(#1){&\lmargin(#1)\tag(#1)\cr}  
\def\tag(#1){\lmargin(#1)({\rm \actualnumber}.\number\EQNcount)
 \NEWDEF e,#1,(\actualnumber.\number\EQNcount)
\global\advance\EQNcount by 1
}
\def\SECT(#1)#2\par{\lmargin(#1)\SECTION#2\par
\NEWDEF s,#1,{\actualnumber}
}
\def\SUBSECT(#1)#2\par{\lmargin(#1)
\SUBSECTION#2\par 
\NEWDEF s,#1,{\actualnumber.\number\SUBSECTIONcount}
}
\def\CLAIM #1(#2) #3\par{
\vskip.1in\medbreak\noindent
{\lmargin(#2)\bf #1\ \actualnumber.\number\CLAIMcount.} {\sl #3}\par
\NEWDEF c,#2,{#1\ \actualnumber.\number\CLAIMcount}
\global\advance\CLAIMcount by 1
\ifdim\lastskip<\medskipamount
\removelastskip\penalty55\medskip\fi}
\def\CLAIMNONR #1(#2) #3\par{
\vskip.1in\medbreak\noindent
{\lmargin(#2)\bf #1.} {\sl #3}\par
\NEWDEF c,#2,{#1}
\global\advance\CLAIMcount by 1
\ifdim\lastskip<\medskipamount
\removelastskip\penalty55\medskip\fi}
\def\SECTION#1\par{\vskip0pt plus.3\vsize\penalty-75
    \vskip0pt plus -.3\vsize
    \global\advance\SECTIONcount by 1
    \beforesectionskip\noindent
{\sectionsize\sectiontype \actualnumber.\ #1}
    \EQNcount=1
    \CLAIMcount=1
    \SUBSECTIONcount=1
    \nobreak\sectionskip\noindent}
\def\SECTIONNONR#1\par{\vskip0pt plus.3\vsize\penalty-75
    \vskip0pt plus -.3\vsize
    \global\advance\SECTIONcount by 1
    \beforesectionskip\noindent
{\sectionsize\sectiontype  #1}
     \EQNcount=1
     \CLAIMcount=1
     \SUBSECTIONcount=1
     \nobreak\sectionskip\noindent}
\def\SUBSECTION#1\par{\vskip0pt plus.2\vsize\penalty-75%
    \vskip0pt plus -.2\vsize%
    \beforesectionskip\noindent%
{\subsectionsize\subsectiontype \actualnumber.\number\SUBSECTIONcount.\ #1}
    \global\advance\SUBSECTIONcount by 1
    \nobreak\sectionskip\noindent}
\def\SUBSECTIONNONR#1\par{\vskip0pt plus.2\vsize\penalty-75
    \vskip0pt plus -.2\vsize
\beforesectionskip\noindent
{\subsectionsize\subsectiontype #1}
    \nobreak\sectionskip\noindent\noindent}
\def\ifundefined#1{\expandafter\ifx\csname#1\endcsname\relax}
\def\equ(#1){\ifundefined{e#1}$\spadesuit$#1\else\csname e#1\endcsname\fi}
\def\clm(#1){\ifundefined{c#1}$\spadesuit$#1\else\csname c#1\endcsname\fi}
\def\sec(#1){\ifundefined{s#1}$\spadesuit$#1
\else Section \csname s#1\endcsname\fi}
\def\fig(#1){\ifundefined{fig#1}$\spadesuit$#1\else\csname fig#1\endcsname\fi}
\let\endarg=\par
\def\finish{\def\endarg{\par\endgroup}}
\def\start{\endarg\begingroup}

 \def\beginFROM{\start\parskip=0pt\vskip\baselineskip
\def\finish{\def\endarg{\egroup\par\endgroup}}
  \vbox\bgroup\obeylines\eightpoint\em\finish}

\def\ABSTRACT#1\par{
\vskip 1in {\noindent\sectionsize\sectiontype Abstract.} #1 \par}

\def\TODAY{\number\day~\ifcase\month\or January \or February \or March \or
April \or May \or June
\or July \or August \or September \or October \or November \or December \fi
\number\year\timecount=\number\time
\divide\timecount by 60
}
\newcount\timecount
\def\DRAFT{\def\lmargin(##1){\strut\vadjust{\kern-\strutdepth
\vtop to \strutdepth{
\baselineskip\strutdepth\vss\rlap{\kern-1.2 truecm\eightpoint{##1}}}}}
\font\footfont=cmti7
\footline={{\footfont \hfil File:\jobname, \TODAY,  \number\timecount h}}
}
\newbox\strutboxJPE
\setbox\strutboxJPE=\hbox{\strut}
\def\subitem#1#2\par{\vskip\baselineskip\vskip-\ht\strutboxJPE{\item{#1}#2}}
\gdef\strutdepth{\dp\strutbox}
\def\lmargin(#1){}
\def\period{\unskip.\spacefactor3000 { }}
%
%
\newbox\noboxJPE
\newbox\byboxJPE
\newbox\paperboxJPE
\newbox\yrboxJPE
\newbox\jourboxJPE
\newbox\pagesboxJPE
\newbox\volboxJPE
\newbox\preprintboxJPE
\newbox\toappearboxJPE
\newbox\bookboxJPE
\newbox\bybookboxJPE
\newbox\publisherboxJPE
\newbox\inprintboxJPE
\def\refclearJPE{
   \setbox\noboxJPE=\null             \gdef\isnoJPE{F}
   \setbox\byboxJPE=\null             \gdef\isbyJPE{F}
   \setbox\paperboxJPE=\null          \gdef\ispaperJPE{F}
   \setbox\yrboxJPE=\null             \gdef\isyrJPE{F}
   \setbox\jourboxJPE=\null           \gdef\isjourJPE{F}
   \setbox\pagesboxJPE=\null          \gdef\ispagesJPE{F}
   \setbox\volboxJPE=\null            \gdef\isvolJPE{F}
   \setbox\preprintboxJPE=\null       \gdef\ispreprintJPE{F}
   \setbox\toappearboxJPE=\null       \gdef\istoappearJPE{F}
   \setbox\inprintboxJPE=\null        \gdef\isinprintJPE{F}
   \setbox\bookboxJPE=\null           \gdef\isbookJPE{F}  \gdef\isinbookJPE{F}
     
   \setbox\bybookboxJPE=\null         \gdef\isbybookJPE{F}
   \setbox\publisherboxJPE=\null      \gdef\ispublisherJPE{F}
     
}
\def\widestlabel#1{\setbox0=\hbox{#1\enspace}\refindent=\wd0\relax}
\def\ref{\refclearJPE\bgroup}
\def\no   {\egroup\gdef\isnoJPE{T}\setbox\noboxJPE=\hbox\bgroup}
\def\by   {\egroup\gdef\isbyJPE{T}\setbox\byboxJPE=\hbox\bgroup}
\def\paper{\egroup\gdef\ispaperJPE{T}\setbox\paperboxJPE=\hbox\bgroup}
\def\yr{\egroup\gdef\isyrJPE{T}\setbox\yrboxJPE=\hbox\bgroup}
\def\jour{\egroup\gdef\isjourJPE{T}\setbox\jourboxJPE=\hbox\bgroup}
\def\pages{\egroup\gdef\ispagesJPE{T}\setbox\pagesboxJPE=\hbox\bgroup}
\def\vol{\egroup\gdef\isvolJPE{T}\setbox\volboxJPE=\hbox\bgroup\bf}
\def\preprint{\egroup\gdef
\ispreprintJPE{T}\setbox\preprintboxJPE=\hbox\bgroup}
\def\toappear{\egroup\gdef
\istoappearJPE{T}\setbox\toappearboxJPE=\hbox\bgroup}
\def\inprint{\egroup\gdef
\isinprintJPE{T}\setbox\inprintboxJPE=\hbox\bgroup}
\def\book{\egroup\gdef\isbookJPE{T}\setbox\bookboxJPE=\hbox\bgroup\em}
\def\publisher{\egroup\gdef
\ispublisherJPE{T}\setbox\publisherboxJPE=\hbox\bgroup}
\def\inbook{\egroup\gdef\isinbookJPE{T}\setbox\bookboxJPE=\hbox\bgroup\em}
\def\bybook{\egroup\gdef\isbybookJPE{T}\setbox\bybookboxJPE=\hbox\bgroup}
\newdimen\refindent
\refindent=5em
\def\endref{\egroup \sfcode`.=1000
 \if T\isnoJPE
 \hangindent\refindent\hangafter=1
      \noindent\hbox to\refindent{[\unhbox\noboxJPE\unskip]\hss}\ignorespaces
     \else  \noindent    \fi
 \if T\isbyJPE    \unhbox\byboxJPE\unskip: \fi
 \if T\ispaperJPE \unhbox\paperboxJPE\unskip\period \fi
 \if T\isbookJPE {\it\unhbox\bookboxJPE\unskip}\if T\ispublisherJPE, \else.
\fi\fi
 \if T\isinbookJPE In {\it\unhbox\bookboxJPE\unskip}\if T\isbybookJPE,
\else\period \fi\fi
 \if T\isbybookJPE  (\unhbox\bybookboxJPE\unskip)\period \fi
 \if T\ispublisherJPE \unhbox\publisherboxJPE\unskip \if T\isjourJPE, \else\if
T\isyrJPE \  \else\period \fi\fi\fi
 \if T\istoappearJPE (To appear)\period \fi
 \if T\ispreprintJPE Pre\-print\period \fi
 \if T\isjourJPE    \unhbox\jourboxJPE\unskip\ \fi
 \if T\isvolJPE     \unhbox\volboxJPE\unskip\if T\ispagesJPE, \else\ \fi\fi
 \if T\ispagesJPE   \unhbox\pagesboxJPE\unskip\  \fi
 \if T\isyrJPE      (\unhbox\yrboxJPE\unskip)\period \fi
 \if T\isinprintJPE (in print)\period \fi
\filbreak
}
\def\hexnumber#1{\ifcase#1 0\or1\or2\or3\or4\or5\or6\or7\or8\or9\or
 A\or B\or C\or D\or E\or F\fi}
\textfont\msbfam=\tenmsb
\scriptfont\msbfam=\sevenmsb
\scriptscriptfont\msbfam=\fivemsb
\mathchardef\varkappa="0\hexnumber\msbfam7B
\newcount\FIGUREcount \FIGUREcount=0
\newdimen\figcenter
\def\definefigure#1{\global\advance\FIGUREcount by 1%
\NEWDEF fig,#1,{Fig.\ \number\FIGUREcount}
\immediate\write16{  FIG \number\FIGUREcount : #1}}
\def\figure #1 #2 #3 #4\cr{\null%
\definefigure{#1}
{\goodbreak\figcenter=\hsize\relax
\advance\figcenter by -#3truecm
\divide\figcenter by 2
\midinsert\vskip #2truecm\noindent\hskip\figcenter
\includegraphics{#1}\vskip 0.8truecm\noindent \vbox{\eightpoint\noindent
{\bf\fig(#1)}: #4}\endinsert}}
\def\figurewithtex #1 #2 #3 #4 #5\cr{\null%
\definefigure{#1}
{\goodbreak\figcenter=\hsize\relax
\advance\figcenter by -#4truecm
\divide\figcenter by 2
\midinsert\vskip #3truecm\noindent\hskip\figcenter
\includegraphics{#1}{\hskip\texpscorrection\input #2 }\vskip 0.8truecm\noindent \vbox{\eightpoint\noindent
{\bf\fig(#1)}: #5}\endinsert}}
\def\figurewithtexplus #1 #2 #3 #4 #5 #6\cr{\null%
\definefigure{#1}
{\goodbreak\figcenter=\hsize\relax
\advance\figcenter by -#4truecm
\divide\figcenter by 2
\midinsert\vskip #3truecm\noindent\hskip\figcenter
\includegraphics{#1}{\hskip\texpscorrection\input #2 }\vskip #5truecm\noindent \vbox{\eightpoint\noindent
{\bf\fig(#1)}: #6}\endinsert}}
\catcode`@=11
\def\footnote#1{\let\@sf\empty 
  \ifhmode\edef\@sf{\spacefactor\the\spacefactor}\/\fi
  #1\@sf\vfootnote{#1}}
\def\vfootnote#1{\insert\footins\bgroup\eightpoint
  \interlinepenalty\interfootnotelinepenalty
  \splittopskip\ht\strutbox 
  \splitmaxdepth\dp\strutbox \floatingpenalty\@MM
  \leftskip\z@skip \rightskip\z@skip \spaceskip\z@skip \xspaceskip\z@skip
  \textindent{#1}\footstrut\futurelet\next\fo@t}
\def\fo@t{\ifcat\bgroup\noexpand\next \let\next\f@@t
  \else\let\next\f@t\fi \next}
\def\f@@t{\bgroup\aftergroup\@foot\let\next}
\def\f@t#1{#1\@foot}
\def\@foot{\strut\egroup}
\def\footstrut{\vbox to\splittopskip{}}
\skip\footins=\bigskipamount 
\count\footins=1000 
\dimen\footins=8in 
\catcode`@=12 

\def\AA{{\script A}}
\def\BB{{\script B}}

\def\OO{{\script O}}

\def\HALF{{\textstyle{1\over 2}}}

\def\QED{\hfill\smallskip
         \line{$\hfill{\vcenter{\vbox{\hrule height 0.2pt
	\hbox{\vrule width 0.2pt height 1.8ex \kern 1.8ex
		\vrule width 0.2pt}
	\hrule height 0.2pt}}}$
               \ \ \ \ \ \ }
         \bigskip}
\def\real{{\bf R}}

\def\complex{{\bf C}}
\def\integer{{\bf Z}}
\def\Re{{\rm Re\,}}
\def\Im{{\rm Im\,}}
\def\PROOF{\medskip\noindent{\bf Proof.\ }}
\def\REMARK{\medskip\noindent{\bf Remark.\ }}
\def\LIKEREMARK#1{\medskip\noindent{\bf #1.\ }}
\normalbaselineskip=5.25mm
\baselineskip=5.25mm
\parskip=10pt
\beforesectionskipamount=24pt plus8pt minus8pt
\sectionskipamount=3pt plus1pt minus1pt
\def\em{\it}
\tenpoint
\null
\normalbaselineskip=12pt
\baselineskip=12pt
\parskip=0pt
\parindent=22.222pt
\beforesectionskipamount=24pt plus0pt minus6pt
\sectionskipamount=7pt plus3pt minus0pt
\overfullrule=0pt
\hfuzz=2pt
\nopagenumbers
\headline={\ifnum\pageno>1 {\hss\tenrm-\ \folio\ -\hss} \else
{\hfill}\fi}
\if T\islinuxolivetti
\font\titlefont=cmbx10 scaled\magstep2

\font\toplinefont=cmr10
\font\pagenumberfont=cmr10
\let\tenpoint=\rm
\else
\font\titlefont=ptmb at 14 pt

\font\toplinefont=cmcsc10
\font\pagenumberfont=ptmb at 10pt
\fi
\newdimen\itemindent\itemindent=1.5em

\def\textindent#1{\indent\llap{#1\enspace}\ignorespaces}
\def\item{\par\noindent
\hangindent\itemindent\hangafter=1\relax
\setitemmark}
\def\setitemindent#1{\setbox0=\hbox{\ignorespaces#1\unskip\enspace}%
\itemindent=\wd0\relax
\message{|\string\setitemindent: Mark width modified to hold
         |`\string#1' plus an \string\enspace\space gap. }%
}
\def\setitemmark#1{\checkitemmark{#1}%
\hbox to\itemindent{\hss#1\enspace}\ignorespaces}
\def\checkitemmark#1{\setbox0=\hbox{\enspace#1}%
\ifdim\wd0>\itemindent
   \message{|\string\item: Your mark `\string#1' is too wide. }%
\fi}
\def\SECTION#1\par{\vskip0pt plus.2\vsize\penalty-75
    \vskip0pt plus -.2\vsize
    \global\advance\SECTIONcount by 1
    \beforesectionskip\noindent
{\sectionsize\sectiontype \actualnumber.\ #1}
    \EQNcount=1
    \CLAIMcount=1
    \SUBSECTIONcount=1
    \nobreak\sectionskip\noindent}

\expandafter\xdef\csname
e*\endcsname{(1.1)}
\expandafter\xdef\csname
e**\endcsname{(1.2)}
\expandafter\xdef\csname
e*1\endcsname{(1.3)}
\expandafter\xdef\csname
e*u\endcsname{(1.4)}
\expandafter\xdef\csname
e*w\endcsname{(1.5)}
\expandafter\xdef\csname
e*RM\endcsname{(1.6)}
\expandafter\xdef\csname
e*tau\endcsname{(1.7)}
\expandafter\xdef\csname
e*kstar\endcsname{(1.8)}
\expandafter\xdef\csname
e*delta\endcsname{(1.9)}
\expandafter\xdef\csname
esampling\endcsname{(1.10)}
\expandafter\xdef\csname
e****\endcsname{(1.11)}
\expandafter\xdef\csname
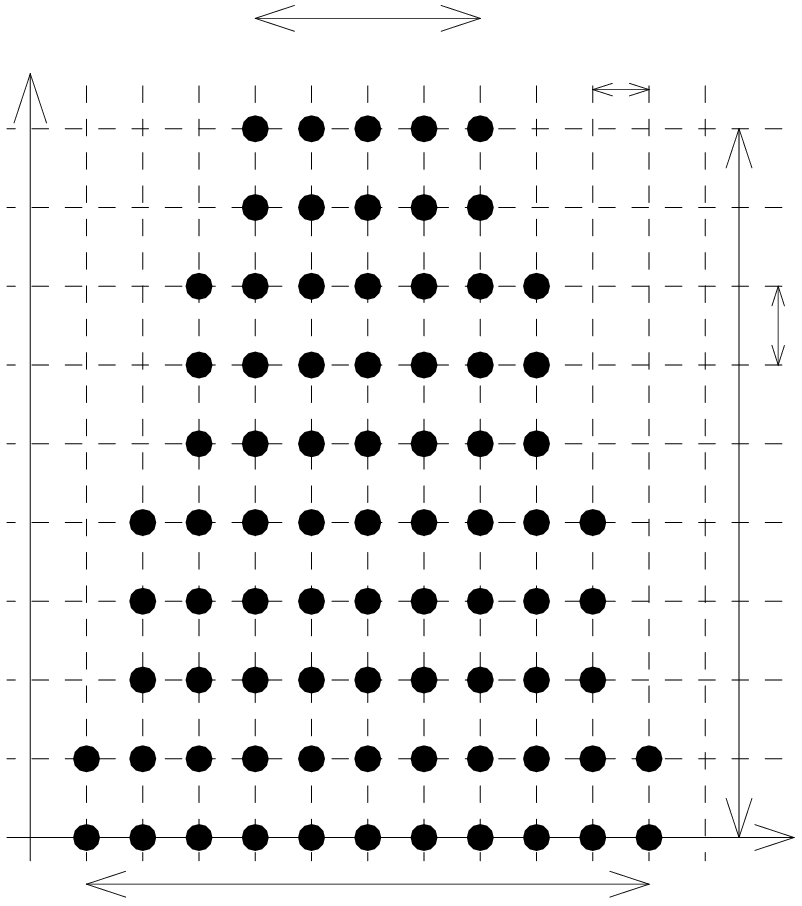\endcsname{Fig.\ 1}
\expandafter\xdef\csname
eelementary\endcsname{(1.12)}
\expandafter\xdef\csname
egdef\endcsname{(2.1)}
\expandafter\xdef\csname
ctoto\endcsname{Lemma\ 2.1}
\expandafter\xdef\csname
elim\endcsname{(2.2)}
\expandafter\xdef\csname
cexist\endcsname{Theorem\ 2.2}
\expandafter\xdef\csname
eaa1\endcsname{(2.3)}
\expandafter\xdef\csname
cgrandir\endcsname{Lemma\ 2.3}
\expandafter\xdef\csname
ehepsilon\endcsname{(3.1)}
\expandafter\xdef\csname
cborne\endcsname{Theorem\ 3.1}
\expandafter\xdef\csname
bBp\endcsname{B^*_{0}}
\expandafter\xdef\csname
ekernelg\endcsname{(4.1)}
\expandafter\xdef\csname
bG\endcsname{B^*_{1}}
\expandafter\xdef\csname
eGnorm\endcsname{(4.2)}
\expandafter\xdef\csname
cgreen\endcsname{Lemma\ 4.1}
\expandafter\xdef\csname
wgym1\endcsname{C_{0}}
\expandafter\xdef\csname
egym1\endcsname{(4.3)}
\expandafter\xdef\csname
wgg\endcsname{C_{1}}
\expandafter\xdef\csname
bHp\endcsname{B^*_{2}}
\expandafter\xdef\csname
ekernel4\endcsname{(4.4)}
\expandafter\xdef\csname
ckernel5\endcsname{Lemma\ 4.2}
\expandafter\xdef\csname
etobound\endcsname{(4.5)}
\expandafter\xdef\csname
etobound1\endcsname{(4.6)}
\expandafter\xdef\csname
etobound2\endcsname{(4.7)}
\expandafter\xdef\csname
bJp\endcsname{B^*_{3}}
\expandafter\xdef\csname
ekernel\endcsname{(4.8)}
\expandafter\xdef\csname
ekernel<\endcsname{(4.9)}
\expandafter\xdef\csname
ckernel\endcsname{Lemma\ 4.3}
\expandafter\xdef\csname
ekernel444\endcsname{(4.10)}
\expandafter\xdef\csname
egym2\endcsname{(4.11)}
\expandafter\xdef\csname
e1\endcsname{(5.1)}
\expandafter\xdef\csname
e2\endcsname{(5.2)}
\expandafter\xdef\csname
e3\endcsname{(5.3)}
\expandafter\xdef\csname
erM\endcsname{(5.4)}
\expandafter\xdef\csname
erdelta\endcsname{(5.5)}
\expandafter\xdef\csname
erMK\endcsname{(5.6)}
\expandafter\xdef\csname
efull\endcsname{(5.7)}
\expandafter\xdef\csname
b01\endcsname{B^*_{4}}
\expandafter\xdef\csname
bk\endcsname{B^*_{5}}
\expandafter\xdef\csname
eboundallk\endcsname{(5.8)}
\expandafter\xdef\csname
eboundlargek\endcsname{(5.9)}
\expandafter\xdef\csname
cdissipative\endcsname{Theorem\ 5.1}
\expandafter\xdef\csname
brho\endcsname{B^*_{6}}
\expandafter\xdef\csname
emore1\endcsname{(5.10)}
\expandafter\xdef\csname
caux1\endcsname{Lemma\ 5.2}
\expandafter\xdef\csname
w0\endcsname{C_{2}}
\expandafter\xdef\csname
wM\endcsname{C_{3}}
\expandafter\xdef\csname
exbound\endcsname{(5.11)}
\expandafter\xdef\csname
w3\endcsname{C_{4}}
\expandafter\xdef\csname
b00\endcsname{B^*_{7}}
\expandafter\xdef\csname
w00\endcsname{C_{5}}
\expandafter\xdef\csname
e77\endcsname{(5.12)}
\expandafter\xdef\csname
bk21\endcsname{B^*_{8}}
\expandafter\xdef\csname
eend\endcsname{(5.13)}
\expandafter\xdef\csname
b00<\endcsname{B^*_{9}}
\expandafter\xdef\csname
e77b\endcsname{(5.14)}
\expandafter\xdef\csname
eendb\endcsname{(5.15)}
\expandafter\xdef\csname
b52\endcsname{B^*_{10}}
\expandafter\xdef\csname
e51\endcsname{(5.16)}
\expandafter\xdef\csname
b99\endcsname{B^*_{11}}
\expandafter\xdef\csname
e52\endcsname{(5.17)}
\expandafter\xdef\csname
bk211\endcsname{B^*_{12}}
\expandafter\xdef\csname
eend99\endcsname{(5.18)}
\expandafter\xdef\csname
e6\endcsname{(6.1)}
\expandafter\xdef\csname
ess\endcsname{(6.2)}
\expandafter\xdef\csname
e4\endcsname{(6.3)}
\expandafter\xdef\csname
bA\endcsname{B^*_{13}}
\expandafter\xdef\csname
bB1\endcsname{B^*_{14}}
\expandafter\xdef\csname
bB\endcsname{B^*_{15}}
\expandafter\xdef\csname
ekprim\endcsname{(6.4)}
\expandafter\xdef\csname
cforward\endcsname{Theorem\ 6.1}
\expandafter\xdef\csname
bcor1\endcsname{B^*_{16}}
\expandafter\xdef\csname
bcor2\endcsname{B^*_{17}}
\expandafter\xdef\csname
ecor1\endcsname{(6.5)}
\expandafter\xdef\csname
ccor\endcsname{Corollary\ 6.2}
\expandafter\xdef\csname
bS\endcsname{B^*_{18}}
\expandafter\xdef\csname
eKnorm4\endcsname{(6.6)}
\expandafter\xdef\csname
ckernel3\endcsname{Lemma\ 6.3}
\expandafter\xdef\csname
bU\endcsname{B^*_{19}}
\expandafter\xdef\csname
bU1\endcsname{B^*_{20}}
\expandafter\xdef\csname
eboundless\endcsname{(6.7)}
\expandafter\xdef\csname
cboundless\endcsname{Lemma\ 6.4}
\expandafter\xdef\csname
efprime\endcsname{(6.8)}
\expandafter\xdef\csname
ef\endcsname{(6.9)}
\expandafter\xdef\csname
csampling\endcsname{Theorem\ 6.5}
\expandafter\xdef\csname
w23\endcsname{C_{6}}
\expandafter\xdef\csname
w24\endcsname{C_{7}}
\expandafter\xdef\csname
ctrick\endcsname{Proposition\ 6.6}
\expandafter\xdef\csname
ewmoredelta\endcsname{(6.10)}
\expandafter\xdef\csname
eksum\endcsname{(6.11)}
\expandafter\xdef\csname
wDss\endcsname{C_{8}}
\expandafter\xdef\csname
bkkk\endcsname{B^*_{21}}
\expandafter\xdef\csname
edefk\endcsname{(6.12)}
\expandafter\xdef\csname
enext1\endcsname{(6.13)}
\expandafter\xdef\csname
enext3\endcsname{(6.14)}
\expandafter\xdef\csname
ewless2a\endcsname{(6.15)}
\expandafter\xdef\csname
bR\endcsname{B^*_{22}}
\expandafter\xdef\csname
ewless2\endcsname{(6.16)}
\expandafter\xdef\csname
endef\endcsname{(6.17)}
\expandafter\xdef\csname
enext5\endcsname{(6.18)}
\expandafter\xdef\csname
bE\endcsname{B^*_{23}}
\expandafter\xdef\csname
bF\endcsname{B^*_{24}}
\expandafter\xdef\csname
bGGG\endcsname{B^*_{25}}
\expandafter\xdef\csname
eallsample\endcsname{(6.19)}
\expandafter\xdef\csname
esuper\endcsname{(6.20)}
\expandafter\xdef\csname
csss\endcsname{Theorem\ 6.7}
\expandafter\xdef\csname
wtime\endcsname{C_{9}}
\expandafter\xdef\csname
e99\endcsname{(6.21)}
\expandafter\xdef\csname
wlast\endcsname{C_{10}}
\expandafter\xdef\csname
wverylast\endcsname{C_{11}}
\expandafter\xdef\csname
eenfin2\endcsname{(6.22)}
\expandafter\xdef\csname
eexper\endcsname{(7.1)}
\expandafter\xdef\csname
ess2\endcsname{(7.2)}
\expandafter\xdef\csname
eexper2\endcsname{(7.3)}
\expandafter\xdef\csname
cexperiment\endcsname{Theorem\ 7.1}
\let\truett=\tt
\fontdimen3\tentt=2pt\fontdimen4\tentt=2pt
\def\tt{\hfill\break\null\kern -2truecm\truett **** }
\def\d{{\rm d}}

\def\GG{{\script G}}
\def\L{{\rm L}}

\let\phi=\varphi
\let\epsilon=\varepsilon
\let\theta=\vartheta

\def\wqeps{{\cal W}_{Q}^{\epsilon}}
\def\attra{{\cal A}}
\def\nustar{\nu_*}
\def\Dstar{D_*}

\def\dup{d_{\rm up}}
\def\Linfty{{\rm L}^\infty }
\def\Dsym{D_{\rm sym}}
\def\Rsym{M_{\rm sym}}

\def\M{M_*}
\def\Mnull{\M}
\def\Q{Q_*}

\def\factor{}
\def\quot{\bigl({\Dstar\over \nustar}\bigr )}

\setitemindent{iii)}
\newcount\Ccount \Ccount=0
\def\C(#1){\lmargin(#1)C_{\number\Ccount}
\NEWDEF w,#1,C_{\number\Ccount}
 \global\advance\Ccount by 1
}
\def\CX(#1){\ifundefined{w#1}\spadesuit#1\else\csname w#1\endcsname\fi}
\newcount\Bcount \Bcount=0
\def\B(#1){\lmargin(#1)B^*_{\number\Bcount}
\NEWDEF b,#1,B^*_{\number\Bcount}
 \global\advance\Bcount by 1
}
\def\BX(#1){\ifundefined{b#1}\spadesuit#1\else\csname b#1\endcsname\fi}
{\titlefont{\centerline{The Definition and Measurement of the }}}
\vskip 0.5truecm
{\titlefont{\centerline{{Topological
Entropy per Unit Volume  }}}
\vskip 0.5truecm
{\titlefont{\centerline{in Parabolic PDE's}}}
\vskip 0.5truecm
{\it{\centerline{ P. Collet${}^{1}$ and J.-P. Eckmann${}^{2,3}$}}}
\vskip 0.3truecm
{\eightpoint
\centerline{${}^1$Centre de Physique Th\'eorique, Laboratoire CNRS UMR
7644,
Ecole Polytechnique, F-91128 Palaiseau Cedex, France}
\centerline{${}^2$D\'ept.~de Physique Th\'eorique, Universit\'e de Gen\`eve,
CH-1211 Gen\`eve 4, Switzerland}
\centerline{${}^3$Section de Math\'ematiques, Universit\'e de Gen\`eve,
CH-1211 Gen\`eve 4, Switzerland}
}}
\vskip 0.5truecm\headline
{\ifnum\pageno>1 {\toplinefont Topological Entropy per Unit Volume}
\hfill{\pagenumberfont\folio}\fi}
\vskip 3cm
{\eightpoint\narrower\baselineskip 11pt
\LIKEREMARK{Abstract} We define the topological entropy per unit
volume in parabolic PDE's such as the complex Ginzburg-Landau
equation, and show that it exists, and is bounded by the upper
Hausdorff dimension times the maximal expansion rate. We then give a
constructive implementation of a bound on the inertial range of such
equations. Using this bound, we are able to propose a finite sampling
algorithm which allows (in principle) to measure this entropy from
experimental data.
}
\SECTION Introduction

In this paper, we shall deal with a general reaction-diffusion
equation, and we have in mind an $N$-component system in $\real^d$
which is of the form
$$
\partial_t u_i(x,t)\,=\,\sum_{j=1}^N d_{ij} \Delta u_j +
F_i(u_1,\dots,u_N)~,\quad
i=1,\dots,N~,
\EQ(*)
$$
where all quantities are real.
For example, the complex Ginzburg-Landau equation (CGL) is
$$
\partial_t v(x,t)\,=\,
(1+i\alpha ) v''(x,t) + v(x,t) - (1+i\beta ) v(x,t) |v(x,t)|^2~,
\EQ(**)
$$
which clearly can be brought to the form of Eq.\equ(*) by writing
equations for the 2 components $u_1=\Re v$, $u_2=\Im v$.
We shall write Eq.\equ(*) short as
$$
\partial_t u\,=\, D\Delta u + F(u)~.
$$
We state now our assumptions on $D$ and $F$. We let
$\Dsym=(D+D^{\rm T})/2$ denote the symmetric part of $D$, and
we assume that the matrix
$\Dsym$ has spectrum $\{\nu_i\}_{i=1,\dots,N}$
in the open right half-line.
We next define
$$
\nustar\,\equiv\,\min_{i=1,\dots,N} \nu_i~,\qquad
\Dstar\,\equiv\,\|D\|~,
\EQ(*1)
$$
where $\|D\|$ is the norm of the matrix $D$ as a linear map from $\real^N$
(equipped with the $l^2 $ norm) to itself.
Note that $\nustar$ is the minimal dissipation rate in Eq.\equ(*).
Our assumptions on $F$ are somewhat vague, but they are intended to
cover a large variety of possible applications.
We first assume that $F$ is ``globally stabilizing'' in the sense that
there is a constant $\Q $ such that for any initial condition $u(x,0)$
which is
bounded in $\Linfty$ there is a $t<\infty $ such that
$\|u(\cdot,t')\|_\infty \le \Q /2$,
for all $t'>t$. (The factor $\HALF$ is
convenient for later use.) In this sense, $\Q $ is the radius
of a globally invariant set (usually this will be an attracting
set).\footnote{${}^1$}{We shall try to stick to the following
notation: Quantities with a $*$ as an index depend on the parameters
of the Eq.\equ(*), and the constants $C_0$, $C_1,\dots$ do {\em not}
depend on them.}
In the case of the real Ginzburg-Landau equation (Eq.\equ(**) with
$\alpha =\beta =0$), one has $\Q /2=1$ and for the CGL it has been shown
in [C,GV] that $\Q <\infty $
when $d=1$ or $d=2$, and also in dimension $d=3$ for some nontrivial
parameter range of $\alpha $ and $\beta $. 
For many other equations one can derive similar bounds using the
localization techniques of [CE1]. Since this is not the central
issue of our paper, we shall just assume that
$$
\|u_t\|_\infty \,\le\, \Q /2~,
\EQ(*u)
$$
for all
$t$. Here, and in the sequel $u_t(x)=u(x,t)$. Our last general
assumption is a bound on the maximal local expansion rate. Consider
two solutions $u$ and $v$ with $\|u_t\|_\infty $ and $\|v_t\|_\infty $
bounded by $Q_*$ for all $t\ge0$. (This is no loss of generality if we
consider later functions in the ``global attractor'' $\AA$.)
We define $w=u-v$. Then, we assume that the
non-linearity $F$ is such that $w$ satisfies an equation of the form
$$
\partial_t w(x,t)\,=\, D\Delta w(x,t) + M(x,t) w(x,t)~,
\EQ(*w)
$$
where the matrix $M$ has a norm (as a map from $\real^N$ to itself)
bounded by
$$
\|M(x,t)\|\,\le\, \Mnull ~,
\EQ(*RM)
$$
for all $x$ and $t$. If $F$ is a polynomial, such a bound will  follow
automatically from the bound of Eq.\equ(*u). For example, for the CGL,
written in complex notation,
we have
$$
\partial_tw\,=\,(1+i\alpha )\Delta w
+\bigl (1+(1+i\beta )( u\bar v + v\bar v)\bigr )w + \bigl ((1+i\beta
)u^2 \bigr )\bar w~,
$$
so that in this case $M$ is a $2\times2$ matrix whose norm is bounded
by
$$
\|M\|\,\le\,1+3(1+|\beta |)\Q ^2/4~.
$$
Again, many other examples can be
handled in this manner and are left to the imagination of the reader.

Our study of the topological entropy is based on a detailed analysis
of the Eq.\equ(*w), and in particular on
the control of {\em information which is localized in space, in the
conjugate momentum,
and in time}.
The localization in space and time has been developed earlier [CE2]
and used
to prove the existence of the $\epsilon$-entropy per unit volume
of Kolmogorov and
Tikhomirov [KT] for systems such as CGL. Here, we use these estimates to
show the existence of the topological entropy per unit volume. We then
improve the bounds to localize at high frequencies, where the flow
defined by
Eq.\equ(*w) will be seen to be essentially a contraction. Using this
information, we shall then show that the topological entropy per unit
volume can be measured in terms of a {\em discrete} sampling of the signal
$u$ in space and time. The amount of data needed for such an
enterprise is, however, quite prohibitive [ER2] if any reasonable
precision is to be attained. But this is probably unavoidable.
For a similar study in finite volume, see [CJT].

We end this introduction by explaining in more detail how the various
physical scales interact as we bound $w$, since this should be useful
to prove further results for dissipative systems in unbounded domains.
We wish to argue in ``dimensionally correct units'' so that $w$ has
the dimension of the observed fields ({\it e.g.}, a temperature),
$[M]\sim [t]^{-1}$ and $[D]\sim [\ell^2 t^{-1}]$, where $t$ is time and
$\ell $ is length. In the long wavelength limit, diffusion is
inactive, and the growth of $w$ is dominated by $M$. Given the a
priori bound $\Mnull $ on $M$, we shall fix the unit of time to
$$
\tau_*\,\equiv\,{1\over \M }~.
\EQ(*tau)
$$
Recall that $\Dstar$ is the norm of
$D$ and the (dimensionless)
quotient $\Dstar/\nustar$ compares essentially the strongest to the weakest
dissipation rates.
The time $\tau_*$ is the time in which errors can grow at most by
a bounded factor, which depends on $\Dstar/\nustar$,
and we shall see that $1/\tau _*$ is also the natural
sampling rate for the determination of the entropy.
The dissipative
range of the equation \equ(*w) is given by those $k$-values for
which dissipation is guaranteed to exceed the growth, {\it i.e.}, for
$\nustar k^2 \ge \M $. Hence we set the cut-off for the $k$-values to
$$
k_*\,\equiv\,\bigl ({1\over \tau _*\nustar}\bigr )^{1/2}
\cdot F(\Dstar/\nustar)~,
\EQ(*kstar)
$$
where the correcting factor $F$ will be defined in Eq.\equ(defk).
The natural unit of length is almost the inverse of $k_*$:
$$
\delta_* \,\equiv\,\bigl ( \tau_*\nustar\bigr )^{1/2} ~.
\EQ(*delta)
$$
(The correcting factor in Eq.\equ(*kstar) is used in the bounds, but
for a more intuitive understanding the reader should assume $\Dstar=\nustar$.)
In terms of these units, we can now explain our ``sampling bound'' of
\clm(sss):
Consider an $\epsilon >0$, which will be the precision we want to
achieve (up to a factor). Assume that two solutions of Eq.\equ(*),
$u$, and $v$  satisfy the bound
$$
|u( m\delta_*,t-n\tau _*)-v( m\delta_*,t-n\tau _*)| \,\le\,\epsilon
~,
\EQ(sampling)
$$
for all $n$ with $0\le n \le C \log(1/\epsilon )$ and for all $|m|\le
(L + C/\epsilon)/\delta_* $ where $C$ is some universal
constant. There is another universal constant $C'$ such that
Eq.\equ(sampling) implies
$$
|u(x,t)-v(x,t)|\,\le\,C'\epsilon ~,
\EQ(****)
$$
for all $|x|\le L$. Thus, a {\em discrete} sampling at spacings
$\delta_*$ and $\tau _*$ suffices to bound the difference of the
two functions {\em everywhere in $\{ |x|\le L\}$.} In other words,
{\em sampling in the inertial range for a time of order $\tau
_*\log(1/\epsilon )$ suffices to bound the dissipative part of $u-v$ as
a function of the sampling precision}.

\figurewithtex figg.ps fig.tex
10.0
8.0
A symbolic representation of the space-time points which need to be
sampled with differences less than $\epsilon $
to get a bound on the difference of two functions on the
interval labeled $L$ to a precision $\BX(GGG)\epsilon$. See \clm(sss) for a
definition of the constants $\BX(E)$, $\BX(F)$ and $\BX(GGG)$.\cr

This result is in line with our
earlier bound [CE2] where we showed that, expressed in the units of the
present paper, the number of balls of radius $\epsilon $ in $\Linfty
$ needed to
cover the global attractor (when restricted to $|x|\le L$) has a bound
of the 
order of
$$
\exp\bigl ( C(L/\delta_*)^{d-1} \log(1/\epsilon )^2\bigr )\cdot
\exp\bigl ( C(L/\delta_*)^d \log
(1/\epsilon )\bigr )~.
$$

One of the main ingredients of the proof of Eq.\equ(****) is the
``dissipative bound'' of \clm(dissipative) below.
In [CE2], we showed the inequality [CE2, Eq.(6.6)]:
$$
\sup_{|x|\le \lambda - \epsilon ^{-1}} |w(x,t)|\,\le\,
 C \epsilon ~,
\EQ(elementary)
$$
when $0\le t\le \tau_*$, and $\sup_{|x|\le \lambda
}|w(x,0)|\le\epsilon $.
We can improve this bound (slightly) as follows and write it in
natural units (see
Eq.\equ(boundallk) for a precise formulation):
$$
\sup_{|x|\le \lambda  -  \ell} |w(x,\tau_*)|\,\le\,
C\bigl (\sup_{|x|\le \lambda } |w(x,0)|+(\delta_* /\ell)
 \|w(\cdot,0)\|_\infty \bigr )\bigl ({\Dstar\over \nustar})^{(1+d)/2}~.
$$
Our improved bound in this paper exploits the dissipation as follows:
Let $P_>$ denote a localization of the Fourier transform of $w$ to
wave numbers
$k$ satisfying $|k|>k_*$ (see below for a more precise
definition). Then,
$$
\sup_{|x|\le \lambda -\ell} |P_> w(x,\tau _*)|\,\le\,
C \bigl ({M_*\over \nustar k_*^2})^{1-d/4}\bigl (\sup _{|x|\,\le\,\lambda } |w(x,0)|+
(\delta_*/\ell) \|w(\cdot,0)\|_\infty\bigr )\bigl ({\Dstar\over
\nustar})^{(1+d)/2} ~. 
$$
Thus, if $k_*$ is sufficiently large, the high frequency part of $w$
decays.

Our paper is organized as follows: We first show that the topological
entropy per unit volume is defined and is bounded by the expansion
rate.
We then show the dissipative bound mentioned above, and finally show
how it can be used to prove the sampling bound. From this, an estimate
of the topological entropy per unit volume can be derived.

\LIKEREMARK{Acknowledgments}This work was partially supported by the
Fonds National Suisse. Our collaboration was also made possible through
the pleasant atmosphere at the IHES, Bures-sur-Yvette.
\SECTION Existence of the Topological Entropy per Unit Volume

We start  by proving  existence of the topological entropy per unit
volume. This is somewhat similar to the standard proof of existence of
topological entropy (see [AKM]). We define the global attractor
$\AA$ by
$$
\AA(B,\tau )\,=\,\bigcap_{n\ge0}\Phi_{n\tau }(B)~.
\EQ(gdef)
$$
Here, $B$ is a ball in $\Linfty$ and $\Phi_t$ is the semi-flow defined by
the evolution Eq.\equ(*).
It can be shown (see [MS]) that $\AA(B,\tau )$ is invariant and that it does
not depend on the initial ball $B$ (if it is large enough) nor on the
(large enough) time $\tau \ge\tau _0(B)$. Thus, we {\em define} $\AA=
\AA\bigl (B,\tau_0(B) \bigr )$.

For any $\epsilon>0$ and any bounded set $Q$ in $\real^d$, whose
boundary has Lebesgue measure 0,
we define $\wqeps$ as the set of all finite coverings of $\attra$
by open sets in $\Linfty(Q)$
of diameter at most $\epsilon$. Note that by the compactness of
$\AA|_Q$, which follows from the uniform bounds on the gradient by
Ascoli [MS],
such finite coverings exist.

Let $\tau>0$ be a fixed time step, and let $T=n\tau$ with $n\in{\integer}$.
For ${\cal U}\in \wqeps$, we say that two trajectories $A_{1}$ and
$A_{2}$  in $\attra$ are $\cal U$-different before time $T$
if there is at least one $j$, $0\le j\le n$ for which the points $\Phi_{j\tau}(A_{1})$
and  $\Phi_{j\tau}(A_{2})$ do not belong to the same element of $\cal U$.
Let
$$
N_{T,\tau}({\cal U })
$$
be the largest number of trajectories which are pairwise $\cal U$-different
before time $T$ (and considered with time-step $\tau $.) Note that
this number is finite since it is at
most $({\rm Card}~ {\cal U})^{T/\tau}$.
Let
$$
N_{Q,T,\tau,\epsilon}=\inf_{{\cal U}\in \wqeps}
N_{T,\tau}({\cal U })~.
$$
\CLAIM Lemma(toto) Let $Q_{1}$ and $Q_{2}$ be two bounded domains
with boundary of zero
Lebesgue measure and $Q_{1}\cap Q_{2}$ of zero Lebesgue measure.
The functions  $N_{Q,T,\tau,\epsilon}$ satisfy the
following bounds:
\item{i)} $N_{Q,T,\tau,\epsilon}$
is non-increasing in $\epsilon$.
\item{ii)} $N_{Q,T_{1}+T_{2},\tau,\epsilon}
\le N_{Q,T_{1},\tau,\epsilon}~N_{Q,T_{2},\tau,\epsilon}$.
\item{iii)} $N_{Q_{1}\cup Q_{2},T,\tau,\epsilon}
\le N_{Q_{1},T,\tau,\epsilon}~N_{Q_{2},T,\tau,\epsilon}$.

\PROOF i) follows since $\wqeps$ is non-decreasing in $\epsilon$.
For a given ${\cal U}\in \wqeps$, we have easily from the definition
$$
N_{T_{1}+T_{2},\tau}({\cal U})
\,\le\, N_{T_{1},\tau}({\cal U})~N_{T_{2},\tau}({\cal
U})~.
$$
Indeed, if we consider a maximal collection of trajectories
$A_{1}\,,A_{2},\,\cdots,\,A_{N_{T_{1}+T_{2},\tau}({\cal
U})}$, we can collect with $A_{1}$ all the remaining $A_{j}$ ($j>1$)
 whose step
$\tau$ orbit $\cal U$ coincides with that of $A_{1}$ up to $T_{1}$. We
know that since these orbits should differ between $T_{1}$ and
$T_{2}$ their number is at most $N_{T_{2},\tau}({\cal U})$.
We continue with the remaining orbits and this leads to at most
$N_{T_{1},\tau}({\cal U})$ groups.  ii) now follows immediately.

In order to prove iii), we consider ${\cal U}_{1}\in {\cal
W}_{Q_{1}}^{\epsilon}$ and ${\cal U}_{2}\in {\cal W}_{Q_{2}}^{\epsilon}$.
Since we are using the $\Linfty $ norm, an
argument similar to the one above leads to
$$
N_{T,\tau}({\cal U}_{1}\times{\cal U}_{2})
\,\le\, N_{T,\tau}({\cal U}_{1})
~N_{T,\tau}({\cal U}_{2})~.
$$
We also have easily (again because we are using the $\Linfty $ norm)
$$
{\cal W}_{Q_{1}}^{\epsilon}\times {\cal W}_{Q_{2}}^{\epsilon}
\subset {\cal W}_{Q_{1}\cup Q_{2}}^{\epsilon}~.
$$
The claim iii) now follows easily.

\LIKEREMARK{Definition}Henceforth, we shall work with domains which
are cubes, and we denote $Q_L$ the cube of side $L$ centered at the
origin.
\CLAIM Theorem(exist) The following limit exists
$$
h\,=\,\lim_{\epsilon\to0}\lim_{L\to\infty}{1\over L^{d}}
\lim_{T\to\infty}{1\over T}
\log N_{Q_{L},T,\tau,\epsilon}~.
\EQ(lim)
$$
Moreover, $h$ does not depend on $\tau$. It is called the topological
entropy per unit volume of the system.

\PROOF From ii) of the \clm(toto), and the usual subadditivity argument,
we conclude that the following limit exists
$$
h^{(1)}_{Q,\tau,\epsilon}\,=\,
\lim_{T\to\infty}{1\over T}
\log N_{Q,T,\tau,\epsilon}~.
$$
Moreover, from i) it is non-increasing in $\epsilon$ and from iii) it
satisfies
$$
h^{(1)}_{Q_{1}\cup Q_{2},\tau,\epsilon}\,\le\,
h^{(1)}_{Q_{1},\tau,\epsilon}+h^{(1)}_{ Q_{2},\tau,\epsilon}~.
$$
Therefore the following limit exists
$$
h^{(2)}_{\tau,\epsilon}=\lim_{L\to\infty} {1\over L^{d}}
h^{(1)}_{Q_{L},\tau,\epsilon}~.
$$
Moreover, $h^{(2)}_{\tau,\epsilon}$
 is non-increasing in $\epsilon$. This proves that the
limit Eq.\equ(lim) exists.

We now show that it is independent of $\tau$.
As in the treatment of topological entropy for finite dimensional
systems, we start by
giving an equivalent definition. Given a positive number $\zeta$, we say
that two trajectories $A_{1}$ and $A_{2}$ in $\attra$
are $\zeta$-separated
in $Q$ before time $T$ (with time-steps $\tau $) if there exists an integer
$j\in[0,T/\tau ]$ for which
$$
\|\Phi_{j\tau}(A_{1})-\Phi_{j\tau}(A_{2})\|_{\Linfty (Q)}\,\ge\,\zeta~.
$$
We denote by $R_{Q,T,\tau}(\zeta)$ the maximum of the cardinalities of the sets
of trajectories which are pairwise $\zeta$-different before time
$T$.
Since the coverings in ${\cal
W}^{\zeta}_{Q}$ are of diameter less than $\zeta$, one has
$$
R_{Q,T,\tau}(\zeta)\,\le\,  N_{Q,T,\tau,\zeta/2}~.
$$
Let now ${\cal U}$ be a finite covering of $\attra$ by balls of radius
$\zeta$ in $\Linfty (Q)$. Then, if two trajectories differ on $\cal
U$ at some time, their distance is at least $\zeta$. Therefore
$$
N_{Q,T,\tau,2\zeta}\,\le\, R_{Q,T,\tau}(\zeta)~.
$$
These two estimates imply immediately that
$$
\lim_{\epsilon\to0}h^{(2)}_{\tau ,\epsilon}
=\lim_{\epsilon\to0}\lim_{L\to\infty}{1\over L^{d}}\lim_{T\to\infty}
{\log R_{Q_{L},T,\tau}(\epsilon)\over T}~.
\EQ(aa1)
$$
\CLAIM Lemma(grandir) There are numbers $\gamma>0$, $\Gamma>1$ and
$C<\infty $
such that for all $L$, $\epsilon>0$ satisfying
$L>C\epsilon^{-1}$ and for all
$A_1$ and $A_{2}$ in $\attra$ for which
$\|A_{1}-A_{2}\|_{\Linfty (Q_{L})}\le \epsilon$, we have for any
$0\le t\le C^{-1}\epsilon L-1$,
$$
\|\Phi_{t}(A_{1})-\Phi_{t}(A_{2})
\|_{\Linfty (Q_{L-C\epsilon ^{-1}(t+1)})}\,\le\, \Gamma e^{\gamma t}
\epsilon~,
$$
with $\Gamma>1$.

\PROOF See \clm(dissipative), Eq.\equ(boundallk) below.
In fact, the statement was already shown in
[CE2].

Using the above estimate one has easily for $\tau'<\tau$,
$$
R_{Q_{L},T,\tau'}(\epsilon)\,\le\, R_{Q_{L+C\epsilon^{-1}(\tau+1)},T,\tau}
(\epsilon \Gamma^{-1}e^{-\gamma\tau})~,
$$
and
$$
R_{Q_{L},T,\tau}(\epsilon)\,\le\, R_{Q_{L+C\epsilon^{-1}(\tau+1)},T,\tau'}
(\epsilon \Gamma^{-1}e^{-\gamma\tau})~.
$$
It follows now immediately from Eq.\equ(aa1) that
 $\lim_{\epsilon\to0}h^{(2)}_{\epsilon,\tau}$
does not depend on $\tau$.
This concludes the proof of \clm(exist).

\SECTION Upper Bound on the Entropy per Unit Volume

It does not follow from what was said in the previous section that we
have defined a finite number. We now give an upper bound. We first
observe that from the sub-additivity argument we have
$$
\lim_{T\to\infty}{1\over T}
\log R_{Q_{L},T,\tau}(\epsilon)\,=\,\inf_{T}
{1\over T}
\log N_{Q_{L},T,\tau,\epsilon }~.
$$
Therefore, in order to obtain an upper bound, we can fix a $T$ and vary
$L$.

Let $H_{\epsilon}$
denote the $\epsilon$-entropy
per unit volume defined in [KT], (see also [CE2] for the application
to the case at hand).
This means that we first define $N_Q(\epsilon )$ as the minimum number
of balls of radius $\epsilon $ in $\Linfty(Q)$ needed to cover $\AA|_Q$
(the functions on the attractor $\AA$ restricted to $Q$), and then
$$
H_\epsilon \,=\,\lim_{L\to\infty } {\log N_{Q_L}(\epsilon )\over L^d}~.
\EQ(hepsilon)
$$
We define the upper
dimension per unit volume $\dup$ of $\attra$ by
$$
\dup\,=\,\limsup_{\epsilon\to0}{H_{\epsilon}\over\log(1/\epsilon)}~.
$$
Note that it follows from [CE2] that $\dup$ is finite.
It is an open question to prove that the limit exists.
However, we have a bound:
\CLAIM Theorem(borne) The topological entropy per unit volume is
bounded by
$$
h\,\,\le\,\, \dup\,\gamma~,
$$
where $\gamma$ is the constant appearing in \clm(grandir).

\REMARK In terms of the variables which scale like the system, we have
$\gamma= \M $.

\PROOF Let $\eta>0$ be a fixed number. Let $\epsilon_{0}>0$ be small
enough such that for all $\epsilon\in(0,\epsilon_{0})$ we have
$$
{H_{\epsilon}\over\log(1/\epsilon)}\,\le\,
\dup+\eta~.
$$
For a fixed $\epsilon\in(0,\epsilon_{0})$ let $L_{\eta,\epsilon}>0$ be
such that for any $L>L_{\eta,\epsilon}$ we have,
$$
\left|{\log  N_{Q_{L}}(\epsilon)\over L^{d}}-H_{\epsilon}\right|\,\le\, \eta~.
$$

We now give an upper bound on $R_{Q_{L},T,\tau}(\zeta)$.
For $\epsilon\in (0,\epsilon_{0})$, choose a
finite covering $\cal U$ of $\attra$ in $\Linfty (
Q_{L+(T+1)\epsilon^{-1}})$
by balls of radius $\epsilon \Gamma^{-1}e^{-\gamma T}/2$. By [CE2] and the
previous discussion,  we know
that we can find such a covering $\cal U$ of cardinality at most
$$
\exp\bigl ((L+C\epsilon ^{-1}(T+1))^{d}\cdot
(\eta+(\dup+\eta)(\gamma T+\log(2\Gamma\epsilon^{-1})))\bigr )~.
$$
Moreover, from the definition of $\gamma$, if $A_{1}$ and $A_{2}$
belong to the same element of $\cal U$ , it follows from \clm(grandir)
that
$$
\sup_{0\le t\le T}\|\Phi_{t}(A_{1})-\Phi_{t}(A_{2})\|_{\Linfty (Q_{L})}
\,\le\,\epsilon~.
$$
Therefore,
$$
R_{Q,T,\tau}(\epsilon)\,\le\,
\exp\bigl ((L+C\epsilon ^{-1}(T+1))^{d}\cdot
(\eta+(\dup+\eta)(\gamma T+\log(2\Gamma\epsilon^{-1})))\bigr )~.
$$
It follows that
$$
\displaylines{
\lim_{L\to\infty}{1\over L^{d}}\inf_{T}{1\over T}\log
R_{Q_{L},T,\tau}(\epsilon)\,\le\,
\lim_{L\to\infty}{1\over L^{d}}{1\over T}\log
R_{Q_{L},T,\tau}(\epsilon)\cr
\,\le\,
\eta+(\dup+\eta)(\gamma+T^{-1}\log(2\Gamma\epsilon^{-1}))~,\cr
}
$$
and the result follows by letting $T\to\infty$ and then $\eta\to0$.

\SECTION Localization in Momentum Space and Bounds on the Semi-Group

In this section, we deal with some simple bounds on the
semi-group generated by $D\Delta$. We begin by constructing the
localization in momentum space. Let $\chi\ge 0$ be a smooth function
with support in $|k|\le 2$ and which is equal to 1 for $|k|\le 1$.
We also assume $\chi\le1$.
We shall denote $\chi_{k<k_*}=\chi(k/k_*)$ and
$\chi_{k>k_*}=1-\chi(k/k_*)$.
We define the convolution operators $\GG_\tau $, $\GG_{\tau ,>}$ and
$\GG_{\tau ,<}$ by
$$
\eqalign{
\GG_\tau (x)\,&=\,\int \d k\, e^{ikx -Dk^2 \tau }~,\cr
\GG_{\tau,>}(x)\,&=\,\int \d k\, e^{ikx} e^{-D k^2\tau }
\chi_{k>k_*}~,\cr
\GG_{\tau,<}(x)\,&=\,\int \d k\, e^{ikx} e^{-D k^2\tau }
\chi_{k<k_*}\,=\,\GG_\tau(x) -\GG_{\tau ,>}(x)~.\cr
}$$
\LIKEREMARK{Notation}The constants $B^*_0$, $B^*_1,\dots,$ do only
depend on the quotient $\Dstar/\nustar$ (something like the condition
number of the matrix $D$), {\em but not on any other parameters of the
problem, except $d$}. We also recall that the constants
$C_0,\dots,$
are numerical factors which do not depend on any parameters of the
problem except $d$.

We now state and prove various estimates on these kernels.
\CLAIM Lemma(green)  For every $p\ge 0$ there is a constant $\B(Bp)(p)$ such
that for all $\tau>0$ one has the bound
$$
\left \|\int \d k\, e^{ikx} e^{-Dk^2\tau}\right \|
\,\le\,\BX(Bp)(p) {1\over  (\nustar\tau) ^{d/2} \bigl (1+
 {|x|^2\over \Dstar\tau }\bigr )^{p/2}}~.
\EQ(kernelg)
$$
Furthermore, convolution with $\GG_\tau $ is
a well-defined operator on bounded functions
and has norm (as a map from $\Linfty $ to itself) bounded by
$$
\|\GG_{\tau}\|_\infty \,\le\,\B(G) ~,
\EQ(Gnorm)
$$
for some $\BX(G)$.

\PROOF For simplicity, we write the proof for the case of $d=1$, but
with a distinction of the upper and lower bounds ($\Dstar$,
resp. $\nustar$) on the matrix $D$.
For example
$$
\eqalign{
x\int \d k\, e^{ikx}e^{-Dk^2\tau }\,&=\,
\int \d k\,\bigl (-i\partial_k e^{ikx}\bigr )e^{-Dk^2\tau }\cr
\,&=\,-i\int \d k\, e^{ikx}e^{-Dk^2\tau }2Dk\tau ~.\cr
}
$$
Thus, in this case,
$$
\eqalign{
\left |\int \d k\, e^{ikx}e^{-Dk^2\tau }\right |
\,&\le\,|x|^{-1}\int \d k\,e^{-Dk^2\tau }2D|k|\tau\cr
\,&\le\,\C(gym1)\bigl ({\Dstar\tau \over |x|^2}\bigr )^{1/2}{1\over (\nustar\tau )^{1/2}}
\bigl ({\Dstar\over \nustar}\bigr )^{1/2}~.\cr
}
\EQ(gym1)
$$
Using $\Dstar/\nustar\ge1$,
the generalization to arbitrary $d$ and $p$ is easy and is left to the
reader. The second assertion follows by taking $p=d+2$ and
integrating. The reader can also check that
$\BX(G)=\C(gg)(\Dstar/\nustar)^{d+2}$, for some $\CX(gg)$.
\CLAIM Lemma(kernel5) Let $\tau \ge0$.
For all $p\ge0$ there is a constant $\B(Hp)(p)$ such that for all
$z\in \complex^d$
one has a bound
$$
\left \|\int \d k\, e^{ikz} e^{-Dk^2 \tau  } \chi_{k<k_*}\right \|\,\le\,
\BX(Hp)(p) {1\over (k_*^{-2}+\nustar\tau)^{d/2}}{e^{2k_*|\Im z|}\over  
\bigl (1+ {| z|^2\over k_*^{-2}+\Dstar\tau}\bigr )^{p/2}} ~,
\EQ(kernel4)
$$
where $|\Im z|\equiv \sum_{i=1}^d |\Im z_i|$.

\PROOF We get, for $z\in\complex$,
$$
\eqalign{
z^p\int \d k\, e^{ikz}e^{-Dk^2\tau }\chi_{k<k_*}\,&=\,
\int \d k\,\bigl ((-i\partial_k)^p e^{ikz}\bigr )e^{-Dk^2\tau
}\chi(k/k_*)~,\cr
}
$$
and integrating by parts this leads to a finite sum of terms of the form
$$
\OO(1)\int \d k\, e^{ikz}e^{-Dk^2\tau }
(D\tau k)^{n_1}k^{-n_2}k_*^{-n_3}\bigl (\partial_k^{n_3}\chi\bigr )(k/k_*)~,
$$
where $n_1+n_2+n_3=p$, with $n_2\le n_1$. We let 
$f=\partial_k^{n_3}\chi$, and we
write $n_1=s_1+s_2$, $n_2=s_1$, and $n_3=s_3$, where now $s_i\ge0$
and
$p=2s_1+s_2+s_3$.
Thus we need to bound expressions of the form
$$
\int_{|k|\le 2k_*} \d k\, e^{-Dk^2\tau }
(D\tau )^{s_1+s_2}k^{s_2}k_*^{-s_3}f(k)~.
\EQ(tobound)
$$
It will be useful to consider first the case $\nustar \tau <1/k_*^2$.
Then we can bound \equ(tobound) in $d$ dimensions by
$$
\OO(1) \quot^p k_*^d (\nustar\tau )^{s_1+s_2}k_*^{s_2}k_*^{-s_3}\,\le\,
\OO(1) \quot^p k_*^d  k_*^{-2s_1-2s_2}k_*^{s_2}k_*^{-s_3}\,=\,\OO(1)
\quot^p k_*^{d-p}~.
\EQ(tobound1)
$$
In the case when $\nustar\tau\ge1/k_*^2$,
we bound \equ(tobound) by
$$
\eqalign{
\OO(1)\quot^p\int \d k\,& e^{-\nustar k^2\tau }
(\nustar\tau )^{s_1+s_2}|k|^{s_2}(\nustar\tau )^{s_3/2}\cr
\,&\le\,
\OO(1)\quot^p (\nustar\tau)^{-d/2+s_1+s_2-s_2/2+s_3/2}\cr
\,&=\,
\OO(1)\quot^p (\nustar\tau)^{-d/2+p/2}~.
\cr
}
\EQ(tobound2)
$$
Combining \equ(tobound1) and \equ(tobound2), and observing that
$|e^{ikz}|\le e^{2k_*|\Im z|}$ on the support of $\chi_{k<k_*}$, we
conclude the proof of \clm(kernel5).
\CLAIM Lemma(kernel) For every $p\ge 0$ there is a constant $\B(Jp)(p)$ such
that for all $k_*>0$ and all $\tau>0$, one has the following
bounds:
\item{i)}When $\nustar \tau > 1/k_*^2$ one has
$$
\left \| \int \d k\, e^{ikx} e^{-D k^2\tau } \chi_{k>k_*}\right \|
\,\le\, \BX(Jp)(p) {e^{-\nustar k_*^2\tau /2}\over (\nustar\tau) ^{d/2} \bigl (1+
 {|x|^2\over \Dstar\tau }\bigr )^{p/2}}~.
\EQ(kernel)
$$
\item{ii)}When $\nustar \tau \le 1/k_*^2$ one has
$$
\eqalign{
\left \| \int \d k\, e^{ikx} e^{-D k^2\tau } \chi_{k>k_*}\right \|
\,&\le\, \BX(Jp)(p) {e^{-\nustar k_*^2\tau /2}}\cr
&\cdot\Biggl(
{1\over (\nustar\tau) ^{d/2} \bigl (1+
 {|x|^2\over \Dstar\tau }\bigr )^{p/2}}+{1\over k_*^{-d} \bigl (1+
 {k_*^2|x|^2 }\bigr )^{p/2}}\Biggr)~.\cr
}
\EQ(kernel<)
$$

\PROOF Assume first that $\nustar \tau \le 1/k_*^2$. In that case,
$e^{\nustar k_*^2\tau}\le e$, and thus it suffices to produce a bound
without exponential factor. We can write $\chi_{k>k_*}=1
-\chi_{k<k_*}$ and get a bound by combining Eq.\equ(kernelg) with
Eq.\equ(kernel4). This leads to
$$
\eqalign{
\left \|\int \d k\, e^{ikz} e^{-Dk^2 \tau  } \chi_{k>k_*}\right \|
\,&\le\,
e\BX(Bp)(p) {1\over  (\nustar\tau) ^{d/2} \bigl (1+
 {|x|^2\over \Dstar\tau }\bigr )^{d/2}}\bigl ({\Dstar\over
 \nustar}\bigr )^{p/2}\cr
&+e\BX(Hp)(p) {1\over (k_*^{-2}+\nustar\tau)^{d/2}}{1\over  
\bigl (1+ {| x|^2\over k_*^{-2}+\Dstar\tau}\bigr )^{p/2}}\bigl
({\Dstar\over \nustar}\bigr ) ~,
}
\EQ(kernel444)
$$
from which the first assertion follows.
In the case $\nustar \tau \ge 1/k_*^2$, we integrate again by parts
and get to bound an expression of the form (we work again in the case
$d=1$ only):
$$
\eqalign{
x\int \d k\, e^{ikx} &e^{-Dk^2\tau} \chi_{k>k_*}\cr
\,&=\,
-i \int\!\! \d k\,e^{ikx} e^{-Dk^2\tau } 
\left (2Dk(1-\chi(k/k_*)) - k_*^{-1} \bigl (\partial_k\chi\bigr
)(k/k_*)\right )~.\cr
}
\EQ(gym2)
$$
Therefore, we get,
as in \equ(gym1),
$$
\eqalign{
\left |\int \d k\, e^{ikx} e^{-Dk^2\tau} \chi_{k>k_*}\right |
\,&\le\,
|x|^{-1}\OO(1)e^{-\nustar k^2 \tau /2} \int \d k \,e^{-Dk^2\tau/2 }\left
|2Dk\tau\right |\cr
\,&+\,|x|^{-1}\OO(1)e^{-\nustar k^2 \tau /2} \int \d k \,e^{-Dk^2\tau/2 }\left
|k_*^{-1}\chi'(k/k_*)\right |\cr
\,&\le\,
C e^{-\nustar k^2 \tau /2} \bigl ({\Dstar \tau \over |x|^2}\bigr
)^{1/2}
{1\over (\nustar \tau )^{1/2}} \bigl ({\Dstar\over \nustar}\bigr
)^{1/2}\cr
\,&+\,Ce^{-\nustar k^2 \tau /2}{1\over (\nustar \tau
)^{1/2}}{k_*^{-1}\over |x|}\cr
\,&\le\,
C e^{-\nustar k^2 \tau /2} \bigl ({\Dstar \tau \over |x|^2}\bigr
)^{1/2}
{1\over (\nustar \tau )^{1/2}} \bigl ({\Dstar\over \nustar}\bigr
)^{1/2}
~.\cr
}
$$
The generalization to arbitrary $p$ and $d $ is easy and is left to
the reader.

\SECTION The Dissipative Bound

In this section, we consider in detail
the equation\footnote{${}^1$}{For simplicity we assume
isotropy of the diffusion in the $d$ components of the coordinates,
but this requirement could be dropped if desired.}
$$
\dot w(x,t) \,=\, D\Delta w(x,t) + M(x,t) w(x,t)~,
\EQ(1)
$$
where $w$ takes values in $\real^N$.
Our bounds will work in dimensions $d\le3$.
We first state the assumptions of the Introduction in a more precise
form.
We first assume that
$$
\|M(x,t)\| \,\le\, \Mnull ~,
\EQ(2)
$$
for all $x\in\real^d$, $t\in\real_+$. We assume
further that
$$
|w(x,t)|\,\le\, \Q ~,
\EQ(3)
$$
for all $x$, $t$. (This is the reason for the choice of $\Q/2$ in the
bound on $u$.) Here, and in the remainder of the paper, $|\cdot|$ is
the $l^2$-norm
of a vector in $\real^N$.

We will fix the constant $k_*$
only in the next section. 
But we will work here with the following ``comparisons of scales''
which will be essential in the bounds:
$$
\eqalignno{
\tau_*\,&=\,1/M_*~,\NR(rM)
\delta_*^2\,&=\, \Dstar\tau _*~,\NR(rdelta)
k_*^2\,&\ge\,1/(\nustar \tau_*)~.\NR(rMK)
}
$$

We consider next the integral representation of $w_1(\cdot)\equiv w(\cdot,
t=\tau_*)$:
$$
w(x,\tau _*)\,=\,\bigl ( \GG_{\tau_*}\star w\bigr )(x) +
\int_0^{\tau_*}  \d s\, \bigl ( \GG_{\tau_*-s} \star \bigl ( M(\cdot,s)
w(\cdot,s)\bigr )\bigr
)(x)\,\equiv\, w^{(1)}_1 + w^{(2)}_1~.
\EQ(full)
$$
Using the decomposition $\GG_\tau =\GG_{\tau,<}+\GG_{\tau ,>}$,
we split $w^{(1)}_1$ and
$w^{(2)}_1$ into high and low
frequency parts:
$$
w_{1,<}\,=\,  w^{(1)}_{1,<} + w^{(2)}_{1,<}~,\qquad
w_{1,>}\,=\,  w^{(1)}_{1,>} + w^{(2)}_{1,>}~.
$$
Note that $w_<$ has Fourier components in $\{|k|\le 2k_*\}$ and $w_> $
has components in $\{ |k| \ge k_*\}$.
\CLAIM Theorem(dissipative) {\rm(Dissipative bound)} In dimension
$d\le3$, we have the following bounds for some constants
$\B(01)$ and $\B(k)$: 
Upon localizing in space, we have for $0\le \tau \le\tau _*$,
$$
\eqalign{
\sup_{|x|\le \lambda - \ell} |w(x,\tau)|\,&\le\, \BX(01)
 \bigl (\sup_{|x|\le \lambda } |w(x,0)|+(\delta_* /\ell)
 \|w(\cdot,0)\|_\infty \bigr )~. \cr}
\EQ(boundallk)
$$
Localizing in position space and momentum space, we have
$$
\eqalign{
\sup_{|x|\le \lambda -\ell}& |w_>(x,\tau _*)|\cr
\,\le\,
\BX(k)& \left (  e^{-(\nustar k_*^2/\M )/2}+
\bigl ({\M \over \nustar k_*^2}\bigr )^{1-d/4}
\right )\cr
\,&\cdot\, \bigl (\sup
_{|x|\,\le\,\lambda } |w(x,0)|+
(\delta_* /\ell) \|w(\cdot,0)\|_\infty \bigr )~.\cr
}
\EQ(boundlargek)
$$

\PROOF We define, as in [CE1],
a family of space cutoff functions:
Let
$$
\psi_a(x)\,=\,  {Z\over  1+((x-a)^2/\delta_*^2)^{1+d/2}}
\,=\, \psi(x-a)~,
$$
where $Z=\delta_*^{-d}\big /\int \d x (1+x^2)^{-1-d/2}$ is chosen such that
$\int \d x\, \psi(x)=1$.
We start with a bound in $\L^2$:
\CLAIM Lemma(aux1) There is a constant $\B(rho)$ such that the
solution of Eq.\equ(1) satisfies:
$$
\sup_{0\le t\le \tau _*}
\sup_{|a|\le \lambda -\ell }
\int \d x\, \psi_a(x) |w(x,t)|^2\,\le\,
\BX(rho) \bigl ( \sup_{|x|<\lambda} |w(x,0)|^2 + \|w(\cdot,0)\|_\infty ^2
(\delta_*/\ell)^2\bigr )~.
\EQ(more1)
$$

\PROOF Let
$$
X_t\,=\,\int \d x\, \psi_a(x) |w(x,t)|^2
~.
$$
Then we have, from the equations of motion,
$$
\eqalign{
\partial_t X_t \,&=\, \int \d x\, \psi_a  \bigl (w \cdot( D\Delta w+ M w) +(
D\Delta w+ M w)\cdot w\bigr )\cr
\,&=\,
-2\int \d x\, \psi_a (\nabla w)\cdot \Dsym\nabla w
+ 2 \int \d x\, \psi_a  \Rsym |w|^2
-  \int \d x\,  (\nabla\psi_a)\nabla  w \cdot D w
~.
\cr 
}
$$
Here, $\Rsym=(M+M^{\rm T})/2$ and $\|\Rsym\|\le\|M\|\le \Mnull $.
Observe that by our choice of $\psi$ we have $|\nabla\psi|\le \C(0)
\psi/\delta_*$
for some constant $\CX(0)$ independent of $\delta_*$.
Using the definitions \equ(*1) of $\nustar$ and $\Dstar$, we find
$$
\eqalign{
|\partial_t X_t| \,&\le\,
-2\nustar \int \d x\, \psi_a |\nabla w|^2
+ 2 \int \d x\, \psi_a \Mnull  |w|^2
+ 2\CX(0) \int \d x\,  (\Dstar/\delta_*)\psi_a  |w\cdot\nabla w|~.\cr
}
$$
We polarize the term containing $w\cdot\nabla w$ and use the
identity $(\Dstar/\delta _*)^2/\nustar=M_*$.
see Eq.\equ(rdelta),
Then we see that we can find a constant $\C(M)$ such that
$$
\eqalign{
|\partial_t X_t| \,&\le\,
\CX(M) \Mnull  \bigl (1+(\Dstar/\nustar) \bigr )X_t~.
\cr
}
\EQ(xbound)
$$
Coming back to the assumptions of \clm(aux1), we see that when $|a|\le
\lambda -\ell$ we have
$$
\eqalign{
X_0\,=\,\int \d x\, \psi_a |w_0|^2 \,&=\,
\int_{|x|\le \lambda} \d x\, \psi_a |w_0|^2
+\int_{|x|\ge\lambda} \d x\, \psi_a |w_0|^2\cr
\,&\le\,
\sup_{|x|<\lambda} |w(x,0)| ^2+
\C(3)\|w_0\|_\infty^2\bigl ({\delta_*\over \ell}\bigr )^2~. \cr
}
$$
Using this bound on the initial condition, the differential inequality
\equ(xbound), and $\tau _*\M =1$, the
assertion of \clm(aux1) follows with
$\BX(rho)=\exp\bigl (\CX(M)(1+(\Dstar/\nustar))\bigr )$.

We begin the proof of \equ(boundlargek). We deal first with the
bound on
$w_{1,>}^{(2)}(x)$, when $|x|\le\lambda-\ell$.
Consider, for $\tau _*\ge t\ge s \ge 0$, the quantity
$$
\eqalign{
Y_{t,s}\,&=\,
\int \d y\,
\GG_{t-s,>}(x-y) M(y,s) w(y,s)\cr
\,&=\,
\int \d y\,
{\GG_{t-s,>}(x-y)\over \sqrt{\psi(x-y)}}\sqrt{\psi(x-y)}  M(y,s)
w(y,s)~.\cr
}
$$
We consider first the case $\nustar\tau>k_*^{-2}$:
Then, by the Schwarz inequality and Eq.\equ(kernel), we have
$$
\eqalign{
Y_{t,s}^2\,&\le\,
\int \d y\,
{|\GG_{t-s,>}(x-y)|^2\over \psi(x-y)}\Mnull  ^2
\int \d z\,
\psi_x(z) |
w(z,s)|^2~\cr
\,&\le\,
\int \d y\,
{\BX(Jp)(p)^2e^{-\nustar k_*^2(t-s)}\quot^{2p}\over
(\nustar(t-s))^{d} \bigl( 1+ {|x-y|^2\over \Dstar({t-s})}\bigr)^{p}
}
{1+(|x-y|^2/\delta_*^2)^{1+d/2}\over Z}
\cr\,&\cdot\,
\Mnull  ^2
\int \d z\,
\psi_x(z) |
w(z,s)|^2~. \cr
}
$$
Since we deal with
$|x|\le\lambda -\ell$, we get from \equ(more1),
$$
\eqalign{
Y_{t,s}^2
\,&\le\,
\BX(Jp)(p)^2\quot^{2p} Z^{-1}\int \d y\,
{
{e^{-\nustar k_*^2(t-s)}\over \nustar ^d(t-s)^{d}}\bigl (1+{|x-y|^2\over
\Dstar (t-s)}\bigr )^{-p}\bigl (1+{|x-y|^2\over
\delta_*^2}\bigr )^{1+d/2}
}
\cr
\,&\cdot\,\Mnull ^2 \BX(rho) \bigl (K^2 + \Q ^2 \bigl ({\delta_*\over \ell}\bigr
)^2\bigr ) ~,
\cr
}
$$
where $K=\sup_{|x|\le\lambda} |w(x,0)|$ and $\Q \ge\|w(\cdot,0)\|_\infty $.
Note now that by Eq.\equ(rdelta),
$$
\delta_*^2\,=\,\Dstar\tau
_*\,\ge\,\Dstar(t-s)~, 
$$
since $\tau _*\ge t\ge s\ge0$. Therefore,
$$
\eqalign{
Y_{t,s}^2
\,&\le\,
\BX(Jp)(p)^2 \quot^{2p}Z^{-1}\int \d y\,
{
{e^{-\nustar k_*^2(t-s)}\over \nustar ^d(t-s)^{d}}\bigl (1+{|x-y|^2\over
\Dstar (t-s)}\bigr )^{1+d/2-p}
}
\cr
\,&\cdot\,\Mnull ^2 \BX(rho) \bigl (K^2 + \Q ^2 \bigl ({\delta_*\over \ell}\bigr
)^2\bigr ) ~.
\cr
}
$$
Taking $p= d+2$, integrating over $y$, and using again 
$\delta_*^2=\Dstar/\M$, we get, for some constant
$\B(00)$:
$$
\eqalign{
Y_{t,s}^2\,&\le\,
\C(00)\BX(Jp)(p)^2  \quot^{2d+4}\delta_*^{d} {\Dstar^{d/2}(t-s)^{d/2}}
{
e^{-\nustar k_*^2(t-s)}\over \nustar ^d (t-s)^{d}
}
\Mnull ^2 \BX(rho)  \bigl (K + \Q  {\delta_*
\over \ell}\bigr )^2\cr
\,&=\,\bigl (\BX(00)\bigr )^2
\Mnull ^{2-d/2}   \bigl (K + \Q  {\delta_*\over \ell}\bigr )^2\,\,
{e^{-\nustar k_*^2(t-s)}\over \nustar^{d/2} (t-s)^{d/2}}
 ~.}
\EQ(77)
$$
Taking the square root of this bound,
integrating over $s$, (and using at this point the hypothesis
$d\le 3$) we get for all $\tau \in[0,\tau _*]$ and all $|x|\le\lambda-\ell$,
$$
\eqalign{
|w_>^{(2)}(x,\tau )|\,&\le\,\int_0^\tau  \d s\, |Y_{\tau ,s}| \,\le\,
\int_0^\infty \d s\,
\BX(00)
M_{*}^{1-d/4}  (\nustar k_*^2 )^{d/4}  \bigl (K + \Q  {\delta
_*\over \ell}\bigr )
{e^{-\nustar k_*^2s/2}\over  (\nustar k_*^2 s)^{d/4}}\cr
\,&\le\,
\B(k21)
\bigl ({M_{*}  \over \nustar k_*^2}\bigr )^{1-d/4} \bigl ( K+ \Q
{\delta_*\over 
\ell}\bigr )
~.\cr
}
\EQ(end)
$$
This
completes the study of the contribution of $w_{1,>}^{(2)}$ to the bound of
Eq.\equ(boundlargek), when $\nustar\tau>1/k_*^{2}$.

We next deal with the case $\nustar\tau\le 1/k_*^2$. The contribution
corresponding to the first term of Eq.\equ(kernel<) is treated as
before noting that in Eq.\equ(end) we actually integrate over all $\tau
$, and that the inequality $\nustar\tau> 1/k_*^2$ was not used
anywhere except for being able to use \equ(kernel) which is the same
as the first term in \equ(kernel<).
Thus, we consider here only the second term.
Although the bounds are quite similar to the previous case, it might
be better to actually spell them out.
By the Schwarz inequality and Eq.\equ(kernel<), we now bound
$$
\eqalign{
W_{t,s}^2
\,&=\,
\int \d y\,
{\BX(Jp)(p)^2e^{-\nustar k_*^2(t-s)}\quot^{2p}\over
k_*^{-2d} \bigl( 1+ {k_*^2|x-y|^2}\bigr)^{p}
}
{1+(|x-y|^2/\delta_*^2)^{1+d/2}\over Z}
\cr\,&\cdot\,
\Mnull  ^2
\int \d z\,
\psi_x(z) |
w(z,s)|^2~. \cr
}
$$
Since we still deal with
$|x|\le\lambda -\ell$, we get from \equ(more1),
$$
\eqalign{
W_{t,s}^2
\,&\le\,
\BX(Jp)(p)^2 \quot^{2p}Z^{-1}\int \d y\,
{e^{-\nustar k_*^2(t-s)}\over k_*^{-2d}}
\bigl (1+k_*^2|x-y|^2\bigr )^{-p}
\bigl (1+{|x-y|^2\over\delta_*^2}\bigr )^{1+d/2} 
\cr
\,&\cdot\,\Mnull ^2 \BX(rho) 
\bigl (K^2 + \Q ^2 \bigl ({\delta_*\over \ell}\bigr )^2\bigr ) ~.
\cr
}
$$
Note now that by Eq.\equ(rdelta),
$$
\delta_*^2\,=\,\Dstar/M\,\ge\,k_*^{-2}
~. 
$$
This leads to
$$
\eqalign{
W_{t,s}^2
\,&\le\,
(\BX(Jp)(p))^2\quot^{2p} Z^{-1}\int \d y\,
{
{ k_*^{2d}}}\bigl (1+{k_*^2|x-y|^2}\bigr
)^{1+d/2-p}
\cr
\,&\cdot\,\Mnull ^2 \BX(rho) \bigl (K^2 + \Q ^2 \bigl ({\delta_*\over \ell}\bigr
)^2\bigr ) ~.
\cr
}
$$
Taking $p= d+2$ and integrating over $y$ we get this time, for some $\B(00<)$:
$$
\eqalign{
W_{t,s}^2\,&\le\,
\bigl (\BX(00<)\bigr )^2 \delta_*^{d}k_*^{-d} k_*^{2d}
\Mnull ^2   \bigl (K + \Q  {\delta_*
\over \ell}\bigr )^2
\cr
\,&=\,\bigl (\BX(00<)\bigr )^2
\Mnull ^{2}  (\delta_* k_*)^d  \bigl (K + \Q  {\delta_*\over \ell}\bigr )^2
 ~.}
\EQ(77b)
$$
Taking the square root of this bound,
integrating over $s$, and noting that $\nustar k_*^2 \tau \le1$, we get
$$
\eqalign{
\int_0^\tau  \d s\, |W_{\tau ,s}| \,&\le\,
\int_0^\tau  \d s\,
\BX(00<)
\Mnull   (\delta_* k_*)^{d/2}  \bigl (K + \Q  {\delta
_*\over \ell}\bigr )
\cr
\,&=\,\BX(00<) \quot^{d/4}
\bigl ({\Mnull   \over \nustar k_*^2}\bigr )^{1-d/4}\bigl ( K+ \Q
{\delta_*\over 
\ell}\bigr )
~.\cr
}
\EQ(endb)
$$
This completes the bound for $\nustar k_*^2 \tau\le1$.

The contribution from  $w_{1,>}^{(1)}$ is easier
to bound. 
Using the definition of $w_{1,>}^{(1)}$, we split the convolution
integral into the region $|y|\le \lambda $ and its complement. Because
$\nustar\tau_*\ge k_*^{-2}$ by Eq.\equ(rMK) we can use
the bound of Eq.\equ(kernel), 
and we get for the first contribution when $p>d+1$,
$$
\BX(Jp)(p)\left |\int_{|y|\le \lambda } \!\!\! \!\!\!\d y\,\,\,
{e^{-\nustar k_*^2\tau_*/2}\over(\nustar\tau_*)^{d/2} \bigl
(1+{|x-y|^2\over \Dstar \tau_* 
}\bigr )^{p/2}}w(y,0)\right |\,\le\,
\B(52)e^{-\nustar k_*^2\tau_*/2}\sup_{|y|\le\lambda }|w(y,0)|~.\EQ(51) 
$$
For the second term, where $|y|\ge\lambda $, the restriction of the
bound to
$|x|\le\lambda-\ell$ implies $|x-y|\ge\ell$,
and then we get using $\tau_*=1/\M$:
$$
\eqalign{
\BX(Jp)(p)\Biggl |&\int_{|x-y|\ge\ell} \!\!\! \!\!\!\d y\,\,\,\,{e^{-\nustar
k_*^2\tau_*/2}\over(\nustar\tau_*)^{d/2} 
 \bigl (1+{|x-y|^2\over \Dstar \tau_*
}\bigr )^{p/2}}w(y,0)\Biggr |\cr
\,&\le\,
\B(99){(\Dstar \tau_*)^{1/2}\over \ell} (\Dstar\tau _*)^{d/2}{e^{-\nustar k_*^2\tau_*/2}\over(\nustar\tau_*)^{d/2}}\sup_y|w(y,0)|\cr
\,&=\,
\BX(99){\delta_*\over \ell}\bigl ({\Dstar\over \nustar}\bigr
)^{d/2}{e^{-\nustar k_*^2\tau_*/2}}\Q~.\cr 
}\EQ(52)
$$
Combining Eqs.\equ(end)--\equ(52), the inequality \equ(boundlargek) follows.

The proof of Eq.\equ(boundallk) is very similar to the one given above,
and we indicate just the few modifications needed. Instead of the
kernel $\GG_{\tau ,>}$ we now use the kernel $\GG_\tau $, 
and we will call $Z_{t,s}$ the
quantity corresponding to $Y_{t,s}$ but with $\GG_\tau $ replaced by
$\GG_{\tau ,>}$. 
Consider first $\nustar\tau >1/k_*^2$.  Since the bound \equ(kernel)
is of the same type as the bound \equ(kernelg), but without he
exponential factor,
all bounds go through as
before up to the inequality \equ(77), which is replaced by a similar
one, but without the exponential factor.
Taking again the square root and integrating over $s$, we get for $d\le3$,
$$
\eqalign{
|w^{(2)}(x,\tau)|\,&\le\,\int_0^\tau \d s\, |Z_{\tau,s}| \,\le\,
\int_0^{\tau_*} \d s\,
(\BX(00))^2 
\cr\,&\cdot\,
\Mnull ^{1-d/4}  (\nustar k_*^2 )^{d/4}  \bigl (K + \Q  {\delta_*\over
\ell}\bigr )
{1\over  (\nustar k_*^2 s)^{d/4}}\cr
\,&\le\,
\B(k211)
\Mnull ^{1-d/4} 
\bigl ( K+ \Q{\delta_*\over \ell}\bigr )\tau_*^{1-d/4}\cr
\,&=\,
\BX(k211)
\bigl ( K+ \Q{\delta_*\over \ell}\bigr )
~.\cr
}
\EQ(end99)
$$
This bounds the contribution of $w_1^{(2)}$ to Eq.\equ(boundallk) where
$\nustar\tau >1/k_*^2$. In the opposite case, we argue exactly as in
the proof of \equ(boundlargek), since the exponential factor in
\equ(kernel<) was anyway of no use before.
Finally, the
contribution of $w_1^{(1)}$ is bounded in exactly the same way as the
one of $w_{1,>}^{(1)}$, except for the exponential factor, and 
we get the bound \equ(boundallk).
The
proof of \clm(dissipative) is complete.

\SECTION The Sampling

We assume that $w_1=w(\cdot,t=\tau_*)$ satisfies
the ``sampling bound''
$$
|w_1|_{{\delta_*},\lambda} \,\le\, \epsilon ~,
\EQ(6)
$$
where
$$
|f|_{\delta_*,\lambda }\,\equiv\,\sup_{\{n\in
\integer^d~:~\delta_*|n|\le\lambda \}}
|f(n\delta_*)|~,
\EQ(ss)
$$
with
$\epsilon \,\le\,\Q $.
Furthermore, we assume
$$
\sup_{|x|\le\lambda } |w(x,0)|\,\le\,K~.
\EQ(4)
$$
Then we have the
\CLAIM Theorem(forward) Consider the solutions $w(\cdot,t)$ of Eq.\equ(1), and
assume that the bounds \equ(2), \equ(3) hold.
There are constants $\B(A)$, $\B(B1)$, and
$\B(B)$ independent of 
$\epsilon $, $K$, and $\lambda$ such that the following holds:
If the initial condition satisfies \equ(4) and $w(\cdot,\tau _*)$
satisfies Eq.\equ(6) with the definition
Eq.\equ(rdelta) of ${\delta_*} $,
then
$$
\sup_{|x|\le\lambda ' } |w(x,\tau _*)|\,\le\, K'~,
$$
where
$$
K'\,=\, \BX(A) \epsilon  + 3 K/4~,
\EQ(kprim)
$$
and $\lambda '=\lambda -\BX(B1) -\BX(B)/K$.

\REMARK Since $3/4\,<\,1$, Eq.\equ(kprim) shows
that $K'$ is smaller than $K$ as long as $K>4\BX(A)\epsilon $.
(In fact, by choosing different constants---in particular a large
$k_*$---we can achieve any 
ratio $\rho>0$ instead of the $3/4$.)

A corollary of the proof is the following result which can be viewed
as a generalization of the dissipative bounds of \clm(dissipative).
\CLAIM Corollary(cor) In dimension $d\le3$ we have the following
bounds
$$
\eqalign{
\sup_{|x|\le \lambda -\ell} |w(x,\tau _*)|
\,&\le\,
\B(cor1) |w(\cdot,\tau_*)|_{\delta_*,\lambda } + {3\over 4} \sup_{|x|\le
\lambda}|w(x,0)| \cr
\,&\,+\B(cor2){\delta_*\over \ell}\|w(\cdot,0)\|_\infty ~.
\cr
}
\EQ(cor1)
$$

\PROOF Our proof is based on sampling theory for functions in the
Bernstein classes, (see, {\it e.g.,}
[B]).
\LIKEREMARK{Definition}We call $\BB_{S,\sigma}$ the Bernstein class
of entire analytic functions
$f$ bounded by
$$
|f(z)|\,\le\, S e^{\sigma |\Im z|}~.
$$
Our first observation is that $\GG_{\tau ,<}$ maps into a Bernstein class
 $\BB_{B,2k_*}$:
\CLAIM Lemma(kernel3) The convolution operator with kernel
$$
\GG_{\tau,<}(x-y)\,=\,\int \d k\, e^{ik(x-y)} e^{-Dk^2\tau }
\chi_{k<k_*}
$$
is bounded from $\Linfty $ to itself. Furthermore, for $\tau \le\tau
_*$, it is bounded
from $\Linfty$ to the Bernstein class $\BB_{\B(S),2k_*}$, and $\GG_{\tau
,<}\star f$ is an
analytic function of $z$ and one has the bound in $\complex^d$:
$$
\bigl |\bigl (\GG_{\tau,<}\star f\bigr )(z)\bigr | \,\le\,\BX(S)
e^{2k_* |\Im 
z|}\factor\|f\|_\infty ~.
\EQ(Knorm4)
$$

\PROOF This follows at once by taking p=d+2 in \clm(kernel5) and integrating.

Using this result, we can now bound the low-frequency part $w_{1,<}$ of
$w_1(x)=w(x,\tau _*)$:
\CLAIM Lemma(boundless) There is a constant $\B(U)$
such that the term $w_{1,<}$ is bounded by
$$
\|w_{1,<}(z)\|_\infty \,\le\, \B(U1)
 e^{2k_* |\Im z|}\factor\|w(\cdot,0)\|_\infty
\,\le\, \BX(U) e^{2k_* |\Im z|}
~.
\EQ(boundless)
$$

\PROOF We recall the representation
$$
w_{1,<}(\cdot)\,=\,w_<(\cdot,\tau _*)\,=\,
\GG_{\tau _*,<}\star w_0
+\int_0^{\tau _*}
\GG_{\tau _*-s,<}\star \bigl (M(\cdot,s)w(\cdot,s)\bigr )~.
$$
By \clm(kernel3), this is bounded as follows:
$$
|w_<(z,\tau _*)|\,\le\,
e^{2k_*|\Im z|}\bigl (\BX(S)Q_*+\BX(Hp)(p)\,\,\tau _* M_*Q_*\bigr
)\,\equiv\,e^{2k_*|\Im z|} \BX(U)~,
$$
so that Eq.\equ(boundless) is proved.

Now that we have established that 
the function $w_{1,<}$ is entire analytic and
exponentially bounded,
we can use the following sampling result (written for functions in $d=1$):
\CLAIM Theorem(sampling) [B] Assume $f\in \BB_{S,\sigma}$. The
following representations hold:
$$\eqalignno{
{f'(0)\over \sigma}\,&=\, {4\sigma\over \pi^2} \sum_{n=-\infty }^\infty
{(-1)^n\over (2n+1)^2} f\bigl (x_{2n+1}\bigr )~,\NR(fprime)
f(x)\,&=\, f'(0) \sin(\sigma x) + f(0) {\sin( \sigma x)\over \sigma x}
+\sigma x \sin(\sigma x)\sum_{n \ne 0} {(-1)^n\over n\pi(\sigma x -n\pi)} f\bigl (x_{2n}\bigr
)~.\NR(f)
}
$$
Here, $x_n= {n\pi \over  2\sigma}$.

\REMARK In higher dimensions, the sum is over a lattice, and details
are left to the reader.

\PROOF [B]: Theorem 11.5.10 and Eq.(11.3.1). (There is an
obvious dimensional misprint in 11.5.10.)

Note that the sums in Eqs.\equ(fprime) and \equ(f) are absolutely
convergent. We now bound them as follows. Let $N$ be a (large)
integer\footnote{${}^*$}{In this section, $N$ denotes just an integer,
and not the number of components of $u$.}. Then, in the first expression, the sum over the terms with
$|n|>N$
is bounded by $\OO(N^{-1}) \|f\|_\infty $. In the second expression,
we assume for the moment that $x\in(x_{0},x_{2})$. Then the sum over
$|n|>N$ is bounded again by $\OO(N^{-1}) \|f\|_\infty $.

\CLAIM Proposition(trick) Assume that $f\in\BB_{S,\sigma}$. Assume
furthermore that $f$ satisfies the bounds
$$
|f(x_{j})|< \alpha ~ {~~\sl for~~}
|j|\,\le\,
J~,
$$
where $x_j={\pi j/(2\sigma)}$.
Assume furthermore that $\|f\|_\infty \le S'$.
Then, one has
the bound
$$
|f(x)|\,\le\, \C(23) \alpha   + {\C(24)S'\over N}~{~~\sl  for~all~~} 
|x|\le \pi(J-N-1)/(2\sigma)~.
$$
The constants $\CX(23)$ and $\CX(24)$ are independent of the
parameters of the problem. We shall assume $\CX(23)>1$.

\PROOF We first bound $f'(0)$ using Eq.\equ(fprime), and this leads to
$$
\sigma^{-1}|f'(0)|\,\le\, {4\over \pi^2}\alpha   \sum _{|n|<N}{1\over(
2n+1  )^2} + {4\over \pi^2}S'  \sum _{|n|\ge N}{1\over(
2n+1  )^2}~.$$
If $|x|\le J \pi/(2\sigma)  $, then $x\in (x_{2{\hat n}},x_{2{\hat n}+2})$ where ${\hat n}$ is
the integer part of $\sigma x/\pi$. Note that by assumption we know
that $|f(x_{2n})|<\alpha  $ for all $n$ satisfying $|n-{\hat n}|< N$.
Thus, splitting the sum as before and shifting the origin to
${\hat n}\pi/\sigma$,  we get a bound
$$
\eqalign{
\bigl |f\bigl (x-({\hat n}\pi/\sigma)\bigr )\bigr |\,&\le\, {4\over
\pi^2}\alpha   \sum _{|n|<N}{1\over(
2n+1  )^2} + {4\over \pi^2}S'  \sum _{|n|\ge N}{1\over(
2n+1  )^2}\cr
\,&+\, \alpha   + \alpha  \sum_{|n|<N} \left |{\sigma
x\sin(\sigma x)\over n\pi(\sigma x-n\pi)}\right |
+ S' \sum_{|n|\ge N} {1\over  2\pi^2 n^2}~.\cr
}
$$
The assertion follows.

We can complete now the proof of \clm(forward). We first bound $w_{1,>}$.
Setting $\lambda''=\lambda -\ell$ and using the bound
Eq.\equ(boundlargek), we get
$$
\eqalign{
\sup_{|x|\le \lambda''}|w_{1,>}(x)|
\,&~~~\le\,
\BX(k)
\left (e^{-(\nustar k_*^2/\M )/2}+
\bigl ({\M \over \nustar k_*^2}\bigr )^{1-d/4}\right )
\cr
\,&\cdot\, \bigl (\sup
_{|x|\,\le\,\lambda } |w(x,0)|+
(\delta_* /\ell) \|w(\cdot,0)\|_\infty \bigr )~.\cr
}
\EQ(wmoredelta)
$$

We now begin fixing the constants:
Recall that $\Q $, $\M $ and $\delta _*=\Dstar\tau _*$
are given by the parameters of the problem.
We have also assumed, as a hypothesis of \clm(forward), that
$$
\sup_{|x|\le\lambda } |w(x,0)|\,\le\,K~.
$$
We choose $\ell=\delta
_*\Q /K $, so that the last factor in \equ(wmoredelta) is bounded by
$2K$. We next choose $k_*$ so large that
$$
\BX(k)
\left ( e^{-(\nustar k_*^2/\M )/2}+
\bigl ({\M \over \nustar k_*^2}\bigr )^{1-d/4}
\right )
\,\le\,{1\over 8\CX(23)}~.
\EQ(ksum)
$$
More precisely, we let
$$
k_*\,=\,\C(Dss) 16 \CX(23) \bigl (\BX(k)\bigr )^{1/(2-d/2)} 
(M_*/\nustar)^{1/2}\,\equiv\, \B(kkk)(M_*/\nustar)^{1/2}~,
\EQ(defk)
$$
and this is our final choice for $k_*$. Clearly, if $\CX(Dss)$ is
sufficiently large, both terms in the sum \equ(ksum) will contribute
less than $1/(16\CX(23))$. 
\REMARK It is at this point crucial that the construction of the
quantities $B_0^*, B_1^*,\dots,$ did {\em not} depend on $k_*$, since
all these constants depend---as we have said before---only on the
quotient $\Dstar/\nustar$. In particular, for problems where $u$ had
only one component, this would mean that the $B_j^*$ are just pure
numerical factors, since then $\Dstar=\nustar$.
It is also important to note that $k_*$ and $\delta_*$
do {\em not} depend on the quantities $\lambda $
or $K$ which occur in \clm(forward).

Thus, so far, with our choices we conclude from Eq.\equ(wmoredelta) that
$$
\sup_{|x|\le \lambda''}|w_{1,>}(x)|
\,\le\, {1\over 4\CX(23)} K
~.
\EQ(next1)
$$
We next observe that
$w_{1,<}=w_1 -w_{1,>}$ and so \equ(next1) leads to
$$
|w_{1,<}(n{\delta_*} )|\,\le\,
|w_{1}(n{\delta_*} )| + |w_{1,>}(n{\delta_*})|\,\le\,
|w_{1}(n{\delta_*} )|+ {1\over 4\CX(23)} K~,
\EQ(next3)
$$
provided $\delta_*|n|\,\le\,\lambda ''$.

By construction, $w_{1,<}$
has a Fourier transform with support in $|k|\le 2k_*$
and furthermore, by \clm(boundless), it is in $\BB_{S,\sigma}$ with
$S=\BX(U)$ and $\sigma=2k_* $.

We can now apply
\clm(trick) to the function $f=w_{1,<}$, with $S'=S=\BX(U)$. Choosing
$J=[2\sigma\lambda ''/\pi]$ (here, $[\cdot]$ is the integer part) we
conclude that, for
$|x|\le\lambda'\equiv  \lambda'' -\delta_*-\pi(N+1)/(2k_*)$, one has a bound
$$
\eqalign{
|w_{1,<}(x)|\,&\le\,\CX(23)|w_{1,<}|_{\delta_*,\lambda ''} +\CX(24)
N^{-1} \BX(U) ~.\cr}
\EQ(wless2a)
$$
It is useful to introduce $\B(R)= \CX(24)\BX(U)$
Using Eq.\equ(next3), this leads to
$$
\eqalign{
\sup_{|x|\le\lambda'}|w_{1,<}(x)|\,&\le\,\CX(23)\bigl (|w_{1}|_{\delta_*,\lambda ''}+ {1\over 4\CX(23)} K\bigr )+
N^{-1} \BX(R) ~.\cr
}
\EQ(wless2)
$$
We next choose
$$
N\,=\,{1\over K} {4\BX(R)}~.
\EQ(ndef)
$$
Then we get
$$
\sup_{|x|\le\lambda'} |w_{1,<}(x)|\,\le\,
\CX(23)|w_{1}|_{\delta_*,\lambda ''}+ K/2~.
\EQ(next5)
$$
Combining Eqs.\equ(next5) and \equ(next1), we
get
$$
\sup_{|x|\le\lambda'} |w_1(x)|\,\le\,
\sup_{|x|\le\lambda'} |w_{1,<}(x)|
+\sup_{|x|\le\lambda''} |w_{1,>}(x)|\,\le\,
\CX(23)|w_{1}|_{\delta_*,\lambda'}+ {K\over 2}+{K\over 4\CX(23)} ~.
$$
Since we have taken $\CX(23)>1$,
we see that we get finally
$$
\sup_{|x|\le\lambda'} |w_1(x)|\,\le\,
\CX(23)|w_{1}|_{\delta_*,\lambda'}+3K/4~,
$$
and
$$
\eqalign{
\lambda '\,&=\,
\lambda'' - \delta_*-\pi(N+1)/(2k_*)\cr
\,&=\,
\lambda - \delta_*\BX(R) /K-\delta_*-\pi(N+1)/(2k_*) ~.
}
$$
From the definition of $N=\OO(1/K)$ it follows at once that there are
constants $\BX(B1)$ and $\BX(B)$ such that
$$
\lambda '\,\ge\,\lambda -\BX(B1) - \BX(B)/K~.
$$
The proof of \clm(forward) is complete.

\CLAIM Theorem(sss) {\rm(Sampling bound)} Consider the solutions
$w(\cdot,t)$ of Eq.\equ(1), and 
assume that the bounds \equ(2), \equ(3) hold.
There are constants $\B(E)$, $\B(F)$ and $\B(GGG)$ such that if
$$
|w(i\delta_*,t-j\tau_*)|\,\le\,\epsilon ~,
\EQ(allsample)
$$
for all $|i|\le (L+\BX(E)/\epsilon )/\delta_*$ and all $|j|\le
\BX(F)\log(1/\epsilon )$, then
$$
\sup_{|x|\le L} |w(x,t)|\,\le\,\BX(GGG)\epsilon ~.
\EQ(super)
$$

\REMARK It will be seen from the proof that it suffices to sample on a
somewhat smaller, non-rectangular domain for \equ(super) to hold. See
also \fig(figs/figg.ps).

\PROOF We obtain the result of \clm(sss) by iteration of \clm(forward).
With $\epsilon $ given as in the statement of \clm(sss),
let $m$ be the
smallest integer for which
$$
(3/4)^m \Q  \,\le\, \epsilon ~,
$$
and note that then
$$
m\,\le\,\log(\Q /\epsilon )/\log(4/3) +1\,\le\,\C(time)\log(1/\epsilon
)~,
$$
when $\epsilon <\HALF$. The number $m$ will be the number of time steps
needed to achieve a precision $\BX(GGG)\epsilon $ in Eq.\equ(super), where
we define $\BX(GGG)=2\BX(A)$, and $\BX(F)=\CX(time)$.
Let now $K_0= \Q $ and $K_{j+1} = \BX(A)\epsilon + 3K_j/4$, and
let $L_m=L$, and $L_{j-1}=L_j +\delta_* + \BX(B)/ K_{j-1}$.
Assume now
$$
|w(i\delta_*,t-j\tau _*)|\,\le\,\epsilon ~,
\EQ(99)
$$
for all $|i|\le L_{m-i}/\delta_*$ and for all $j=0,\dots,m$.
The reader can check easily that our definitions are made such that
\clm(forward) applies at each time step considered.
Since we have
$$\|w(\cdot,t-j\tau _*)\|_\infty \,\le\,
\Q ~,
$$
for all $j$, we can inductively bound, for $j=1,\dots,m$,
$$
\sup_{|x|\le L_j} |w(x,t-(m-j)\tau _*)|\,\le\,
\CX(23)\epsilon + 3K_{j-1}/4\,=\,K_{j+1}~.
$$
Note now that $K_j=\Q \rho^j + \CX(23)\epsilon (1-\rho^j)/(1-\rho)$,
where $\rho=3/4$ and thus
$$
L_{m-j}\,=\,L + j\delta_* + \BX(B) \sum_{i=1}^j K_{m-i}^{-1}~.
$$
Clearly, there is a $\C(last)>0$ such that
$K_j\ge \CX(last) \rho^j$, so that we get
$$
L_{m-j}\,=\,L + j\delta_* + {\BX(B)\over \CX(last)} \sum_{i=1}^j
{1\over \rho^{m-j}}~.
$$
In particular, we can find some $\C(verylast)$, so that
$$
L_0\,\le\,L+m\delta_* +  {\BX(B)\over \CX(last)} {\rho^{-m}\over
{\rho^{-1}-1}}\,\le\,L + \CX(time)\delta_*\log(1/\epsilon )
+ \CX(verylast)/\epsilon ~.
$$
Note that $L_0$ is the width of the ``earliest'' bound in
\equ(allsample), and we choose $\BX(E)=\CX(time)\delta_*+\CX(verylast)$.
Since $K_m\le \BX(A)\epsilon $ we have shown that the bounds of
Eq.\equ(99) are sufficient to ensure
$$
\sup_{|x|\le L} |w(x,t)|\,\le\,2\BX(A)\epsilon ~.
\EQ(enfin2)
$$
The proof of \clm(sss) is completed.

\SECTION Outlook

In this section, we wish to discuss potential experimental aspects of
our results. These aspects must necessarily rely on a number of
conjectures about the system under consideration, and are similar in
spirit to the discussion found in [ER]. While we have defined
topological entropy in Section 2, we now need to address the question
of entropy relative to an invariant measure $\mu$. We shall call it
$h_\mu$. More precisely, let $f$ be a continuous map of a compact
metric space
and let $\mu$ be an $f$-invariant non-atomic ergodic measure. The entropy
$h_{\mu}(f)$ is determined as follows. For $\epsilon>0$ and an integer
$n>0$ let
$$
V(x,\epsilon,n)\,=\,\big\{ y\;\big|\;d(f^{i}(y),f^{i}(x))<\epsilon\;,\;0\le i<n
\big\}~.
$$
One has (see [BK], [Y]):
$$
h_{\mu}(f)\,=\,-\lim_{\epsilon\to0}\liminf_{n\to\infty}
{1\over n}\log\mu(V(x,\epsilon,n))~.
$$
This is not a very convenient expression for explicit computations. As for
the case of finite dimension (see [GP]) one can try to determine instead a
correlation entropy. In our case we would also like to include the dependence
on the size of the window in which the system is observed. One is
naturally lead to the following definition of correlation entropy per
unit length $K_2(\mu)$---and unit time---(see [GP, ER])
for a measure $\mu$ which is space and time
invariant and ergodic.
The definition of this quantity is: 
$$
\eqalign{
K_2(\mu)\,&=\,-{1\over \tau }\lim_{\epsilon\to0}\lim_{L\to\infty}{1\over L^{d}}\lim_{n\to\infty}
{1\over n}\cr
&~~~
\log\left(\lim_{N\to\infty}{1\over N^{2}}
\sum_{j,k=0}^{N-1}\prod_{i=0}^{n-1}
\Theta\bigg(\epsilon-\|\Phi_{(i+j)\tau}(A)
-\Phi_{(i+k)\tau}(A)\|_{\Linfty(Q_{L})}\bigg)\right)~,}
\EQ(exper)
$$
where $\Theta$ is the Heaviside function, and $\Phi_t$ is the flow of the
dynamics.

It is an open question to prove that the limits in the above expression
exist for $\mu$ almost every $A$ (except the limit over $\epsilon$ since
the quantity is increasing). 
Furthermore, the result should also be independent of
the time step $\tau$. 
But let us assume that these limits exist.
In that case our sampling bound \clm(sss) gives us a constructive handle on
computing the r.h.s of Eq.\equ(exper).
We recall the definition
$$
|f|_{\delta,\lambda }\,\equiv\,\sup_{\{n\in
\integer^d~:~\delta|n|\le\lambda \}}
|f(n\delta)|~.
\EQ(ss2)
$$
\CLAIM Theorem(experiment) Assuming the limits in \equ(exper) exist for
$\mu$-almost every $A$, we have for every $\delta\in (0,\delta _*]$
and every $\tau\in (0,\tau_*]$:
$$
\eqalign{
K_2(\mu)\,&=\,-{1\over \tau }\lim_{\epsilon\to0}\lim_{L\to\infty}{1\over L^{d}}\lim_{n\to\infty}
{1\over n}\cr
&~~~
\log\left(\lim_{N\to\infty}{1\over N^{2}}
\sum_{j,k=0}^{N-1}\prod_{i=0}^{n-1}
\Theta\bigg(\epsilon-|\Phi_{(i+j)\tau}(A)
-\Phi_{(i+k)\tau}(A)|_{\delta ,L}\bigg)\right)~.}
\EQ(exper2)
$$

\REMARK Note that the sum $\sum_{j,k=0}^{N-1}\prod_{i=0}^{n-1}\Theta(\dots)$
counts the number of pairs of points in the sample which have a
``distance'' of less than $\epsilon $ in embedding dimension $n$, {\em
where the
distance is measured with the discrete sampling step $\delta $ over
a region $|x|\le L$.} In this sense, our result says that for PDE's on
the infinite line, measuring the $K_2$ entropy can be done by the
usual Grassberger-Procaccia algorithm [GP, ER].
\REMARK If we choose $\delta =\delta _*$ and $\tau =\tau_*$, then the
``sampling error'' is controlled by the constants of the proof of
\clm(sss).
But if
we choose, for example $\tau=\tau_*/2$, the number of time steps---{\it
i.e.,} the factor $\BX(F)$---needed in \clm(sss) will be the same as
if $\M$ had been replaced by 
$2\M$ in the original assumptions. So clearly, the total time during
which one must measure to achieve a given sampling precision can not
be shortened by sampling at shorter intervals. Similar precautions are
necessary for sampling in space.

\PROOF In view of the remark, we give the proof only for the case $\tau
=\tau _*$, $\delta =\delta _*$, in which case the constants
$\BX(E),\dots,$ retain their meaning from earlier parts of the paper.
We obviously have
$$
\displaylines{
{1\over N^{2}}
\sum_{j,k=0}^{N-1}\prod_{i=0}^{n-1}
\Theta\bigg(\epsilon-|\Phi_{(i+j)\tau_*}(A)
-\Phi_{(i+k)\tau_*}(A)|_{\delta_*, L}\bigg)\cr
\,\ge\,
{1\over N^{2}}
\sum_{j,k=0}^{N-1}\prod_{i=0}^{n-1}
\Theta\bigg(\epsilon-\|\Phi_{(i+j)\tau_*}(A)
-\Phi_{(i+k)\tau_*}(A)\|_{\Linfty(Q_{L})}\bigg)~.
\cr
}
$$
On the other hand, applying \clm(sss), we have for
$n>\BX(F)\log(1/\epsilon)+1$ and $L>\BX(E)/\epsilon$:
$$
\eqalign{
\prod_{i=0}^{n-1}
\Theta\bigg(&\epsilon-|\Phi_{(i+j)\tau_*}(A)
-\Phi_{(i+k)\tau_*}(A)|_{\delta_*, L}\bigg)
\cr
\,\le\,
\prod_{i=[\BX(F)\log(1/\epsilon)]}^{n-1}
\Theta\bigg(\BX(GGG)&\epsilon-\|\Phi_{(i+j)\tau_*}(A)
-\Phi_{(i+k)\tau_*}(A)\|_{\Linfty(Q_{L-\BX(E)/\epsilon})}\bigg)\cr
\,\le\,
\prod_{i=0}^{n-1-[\BX(F)\log(1/\epsilon)]}
\Theta\bigg(\BX(GGG)&\epsilon-\|\Phi_{(i+j+[\BX(F)\log(1/\epsilon)])\tau_*}(A)
\cr~~~~~
&~~-\Phi_{(i+k+[\BX(F)\log(1/\epsilon)])\tau_*}(A)
\|_{\Linfty(Q_{L-\BX(E)/\epsilon})}\bigg)~.
}
$$
Therefore,
$$
\eqalign{
{1\over N^{2}}&
\sum_{j,k=0}^{N-1}\prod_{i=0}^{n-1}
\Theta\bigg(\epsilon-|\Phi_{(i+j)\tau_*}(A)
-\Phi_{(i+k)\tau_*}(A)|_{\delta_*, L}\bigg)
\cr
\,&\le\,
{1\over N^{2}}
\sum_{j,k=0}^{N-1}
\prod_{i=0}^{n-1-[\BX(F)\log(1/\epsilon)]}
\Theta\bigg(\BX(GGG)\epsilon-\|\Phi_{(i+j)\tau_*}(A)
-\Phi_{(i+k)\tau_*}(A)\|_{\Linfty(Q_{L-\BX(E)/\epsilon})}\bigg)
\cr
&~~+{1\over N}\OO\big(1+\log(1/\epsilon)\big)^{2}~.
\cr
}
$$
The result now follows by taking the limits $N\to\infty$, $n\to\infty$,
$L\to\infty$ (in that order) for $\epsilon$ fixed, and then letting
$\epsilon$ tend to zero.
\SECTIONNONR References

\widestlabel{[XXX]}
{\eightpoint
\ref
\no AKM
\by Adler, R., Konheim, A.G. and McAndrew, M.H.
\paper Topological Entropy
\jour Trans. Am. Math. Soc.
\pages 390
\vol 114
\yr 1965
\endref
\ref 
\no B
\by Boas, R.P.
\book Entire Functions
\publisher New York: Academic Press
\yr 1954
\endref
\ref
\no BK
\by Brin, M. and Katok, A.
\paper On local entropy
\inbook Geometric Dynamics {\rm (Rio de Janeiro,
1981), 
Lecture Notes in Mathematics} {\bf 1007}
\pages 30--38
\yr 1983
\endref
\ref 
\no CJT
\by Cockburn, B., Jones, D.A. and Titi, E.
\paper Estimating the number of asymptotic degrees of freedom for
nonlinear dissipative systems
\jour Math. Comput.
\vol 66
\pages 1073--1087
\yr 1997
\endref
\ref
\no C
\by Collet, P.
\paper Thermodynamic limit of the Ginzburg-Landau equation
\jour Nonlinearity
\vol 7
\pages 1175--1190
\yr 1994
\endref
\ref
\no CE1
  \by Collet, P. and Eckmann, J.-P.
  \paper The time-dependent amplitude equation for the Swift-Hohenberg problem
  \jour Commun. Math. Phys.
  \vol 132
  \pages 139--153
  \yr 1990
\endref
\ref
\no CE2
  \by Collet, P. and Eckmann, J.-P.
  \paper Extensive properties of the complex Ginzburg-Landau
equation
\preprint
  \yr 1998
\endref
 \ref
\no ER
  \by Eckmann, J.-P. and Ruelle, D.
  \paper Ergodic theory of chaos and strange attractors
  \jour Rev. Mod. Phys.
  \vol 57
  \pages 617--656
  \yr 1985
\endref
\ref
\no ER2
  \by Eckmann, J.-P. and Ruelle, D.
\paper Fundamental limitations for estimating dimensions and Liapunov
exponents in 
dynamical systems
\jour Physica 
\vol D56
\pages 185--187 
\yr 1992
\endref
\ref 
\no GP
\by Grassberger, P. and I. Procaccia
\paper Estimating the Kolmogorov entropy from a chaotic signal
\jour Phys. Rev.
\vol A28
\pages 2591
\yr 1983
\endref
\ref
\no GV
\by Ginibre, J. and G. Velo
\paper The Cauchy problem in local spaces for the complex
  Ginzburg-Landau equation. II: contraction methods
\jour Commun. Math. Phys.
\vol 187
\pages 45--79
\yr 1997
\endref
\ref
\no KT
 \by Kolmogorov, A.N. and Tikhomirov, V.M.
 \paper $\epsilon$-entropy and $\epsilon $-capacity of sets in
 functional spaces\footnote{${}^1$}{The version in this collection is
 more complete than the original paper of Uspekhi Mat. Nauk, {\bf 14},
 3--86 (1959).}
 \inbook Selected Works of A.N Kolmogorov, Vol III
 \bybook Shirayayev, A.N., ed.
 \publisher Dordrecht, Kluver
 \yr 1993
\endref
\ref
\no MS
  \by Mielke, A. and Schneider, G.
  \paper Attractors for modulation equations on unbounded
 domains---existence and comparison
  \jour Nonlinearity
  \vol 8
  \pages 743--768
  \yr 1995
\endref
\ref
\no Y
\by Young, L.S.
\paper Dimension, entropy and Lyapunov exponents
\jour Erg. Th. Dyn. Sys.
\vol 2
\pages 109--124 
\yr 1982
\endref
}
\bye